\documentclass[sigconf, nonacm]{acmart}

\usepackage{graphicx}
\usepackage{textcomp}
\usepackage{xcolor}
\usepackage{multirow}
\usepackage{xspace}
\usepackage{listings}
\usepackage{subfigure}
\usepackage{algorithm}
\usepackage{algpseudocode}
\usepackage{amsmath}
\usepackage{xspace}
\usepackage{bm}

\let\emptyset\varnothing

\newcommand{\sys}{{{GraSorw}\xspace}}
\newcommand{\cache}{{{SGSC}\xspace}}

\newcommand{\tabincell}[2]{\begin{tabular}{@{}#1@{}}#2\end{tabular}}

\newtheorem{myDef}{Definition}
\newtheorem{myTheorem}{Theorem}
\newtheorem*{myProof*}{Proof}

\newcommand\vldbdoi{XX.XX/XXX.XX}

\newcommand\vldbavailabilityurl{https://github.com/DuoLife-QNL/GraSorw}
\newcommand\vldbpagestyle{plain} 

\begin{document}
\title{An I/O-Efficient Disk-based Graph System for Scalable Second-Order Random Walk of Large Graphs}


\author{
  Hongzheng Li{$^{\star}$},~~~Yingxia Shao{$^{\star *}$},~~~Junping Du{$^{\star}$},~~~Bin Cui{$^{\S}$},~~~Lei Chen{$^{\#}$}
}
\affiliation{%
  \institution{
  {$^{\star}$}School of Computer Science, Beijing University of Posts and Telecommunications \\
  {$^{\S}$}School of EECS \& Key Laboratory of High Confidence Software Technologies (MOE), Peking University \\
  {$^{\#}$}Department of Computer Science and Engineering, Hong Kong University of Science and Technology
  }
}
\email{{Ethan_Lee,shaoyx}@bupt.edu.cn, junpingdu@126.com, bin.cui@pku.edu.cn, leichen@cse.ust.hk}
\thanks{*Yingxia Shao is the corresponding author}



\begin{abstract}
Random walk is widely used in many graph analysis tasks, especially the first-order random walk. 
However, as a simplification of real-world problems, the first-order random walk is poor at modeling higher-order structures in the data. Recently, second-order random walk-based applications (e.g., Node2vec, Second-order PageRank) have become attractive. Due to the complexity of the second-order random walk models and memory limitations, it is not scalable to run second-order random walk-based applications on a single machine. Existing disk-based graph systems are only friendly to the first-order random walk models and suffer from expensive disk I/Os when executing the second-order random walks. 
This paper introduces an I/O-efficient disk-based graph system for the scalable second-order random walk of large graphs, called \sys. First, to eliminate massive light vertex I/Os, we develop a bi-block execution engine that converts random I/Os into sequential I/Os by applying a new triangular bi-block scheduling strategy, the bucket-based walk management, and the skewed walk storage. Second, to improve the I/O utilization, we design a learning-based block loading model to leverage the advantages of the full-load and on-demand load methods. Finally, we conducted extensive experiments on six large real datasets as well as several synthetic datasets. The empirical results demonstrate that the end-to-end time cost of popular tasks in \sys~ is reduced by more than one order of magnitude compared to the existing disk-based graph systems. The source code is available at \url{\vldbavailabilityurl}
\end{abstract}

\maketitle

\pagestyle{\vldbpagestyle}


\section{Introduction}
\label{sec:intro}
Random walk has been successfully used in a variety of graph analysis tasks~\cite{Jeh2002,Perozzi2014,Tsitsulin2018,Fogaras2005,Liu2016,Bar-Yossef2000,Henzinger1999,Dong2017,Rahman2019,Peng2021,Yang2021}. Most of the existing tasks adopt first-order random walk models~\cite{Langville2011,Perozzi2014}, which assume that the next vertex of a walk only relies on the information of the current vertex. However, as a simplification of real-world problems, the first-order random walk is poor at retaining historical information. Previous studies~\cite{Wu2016} show that higher-order random walk models can provide better support for graph analysis tasks by selecting the next vertex based on more historical information. 
{Node2vec~\cite{Grover2016} is one of the most successful applications of the second-order random walk model, and for the graph embedding task, it has better performance than DeepWalk~\cite{Perozzi2014}, which uses the first-order random walk model. For the graph proximity measurements, PageRank~\cite{Kumar1998} and SimRank~\cite{Jeh2002} are two popular metrics. CoSimRank~\cite{Rothe2014} is proposed to reduce the computation cost in SimRank. All of these metrics adopts the first-order random walk model. In recent years, Wu et.al.~\cite{Wu2016} proposed second-order random walk-based PageRank and SimRank, and Liao et.al. put forward the second-order CoSimRank~\cite{Liao2019}. They all demonstrated that the second-order approaches achieve better results compared to the standard ones through empirical studies. Second-order random walk-based models are also widely used in community detection tasks, such as overlapping community detection~\cite{Evans2009,Deng2017} and arc-community detection~\cite{Boldi2012}. Moreover, many other interesting applications adopt second-order random walk to model different complex systems. For example, in cloud services, ServiceRank~\cite{Ma2021} and CloudRanger~\cite{Wang2018} apply the second-order random walk to identify the culprit services which are responsible for cloud incidents. For the intelligent transportation systems, R. Besenczi et.al.~\cite{Besenczi2021} introduced second-order random walk-based model on dual graph~\cite{Porta2006} to analysis the traffic flow on urban streets.}


{Nowadays, many real-world graphs occupy hundreds of Gigabytes in CSR format, which exceeds the size of the RAM for most commodity machines.}
Due to the limitation of memory, it is not scalable to run random walk models on large graphs with memory-based frameworks~\cite{Shao2020,Shao2021} in a single machine.
Many general disk-based graph systems~\cite{Kyrola2012,Zhu2015,Maass2017} are proposed to conduct first-order random walks on large graphs. 
They originally partition the whole graph into several blocks, i.e., subgraphs. During execution, these systems load a block into memory, update all the activated vertices and edges in the current block, and repeat this operation until a certain termination condition is satisfied. DrunkardMob~\cite{Kyrola2013} is the first random walk oriented disk-based graph system, which proposes a framework to start millions of random walks simultaneously, and GraphWalker~\cite{Wang2020} follows its idea while using an asynchronous walk update method to reduce the I/O overhead further.

\begin{figure}[t]
\centering
    \subfigure[I/O-percentage]{
        \label{fig:io-percentage}
        \includegraphics[width=0.22\textwidth]{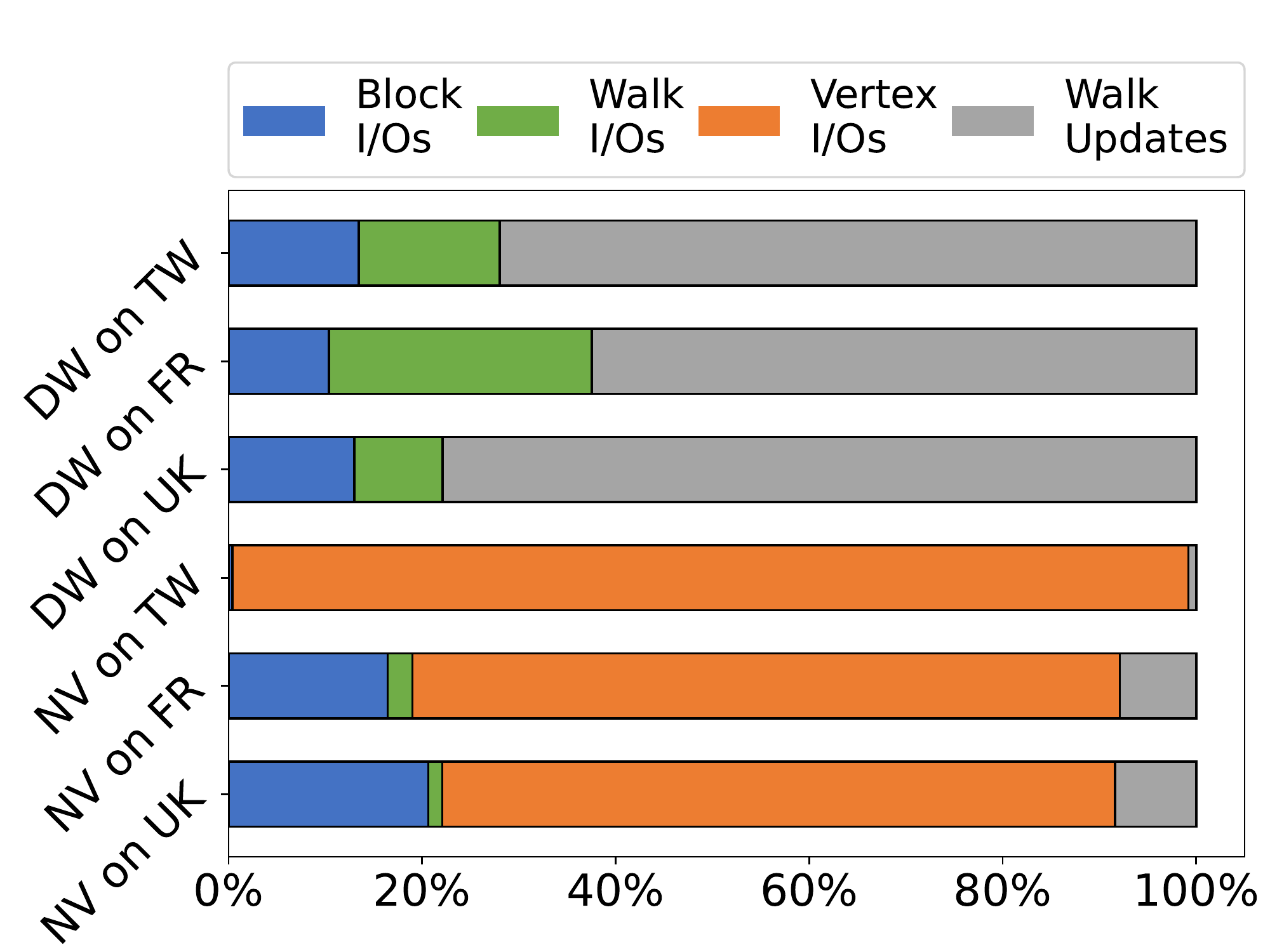}
    }
    \subfigure[Activated vertex Ratio]{
        \label{fig:activated-ratio}
        \includegraphics[width=0.22\textwidth]{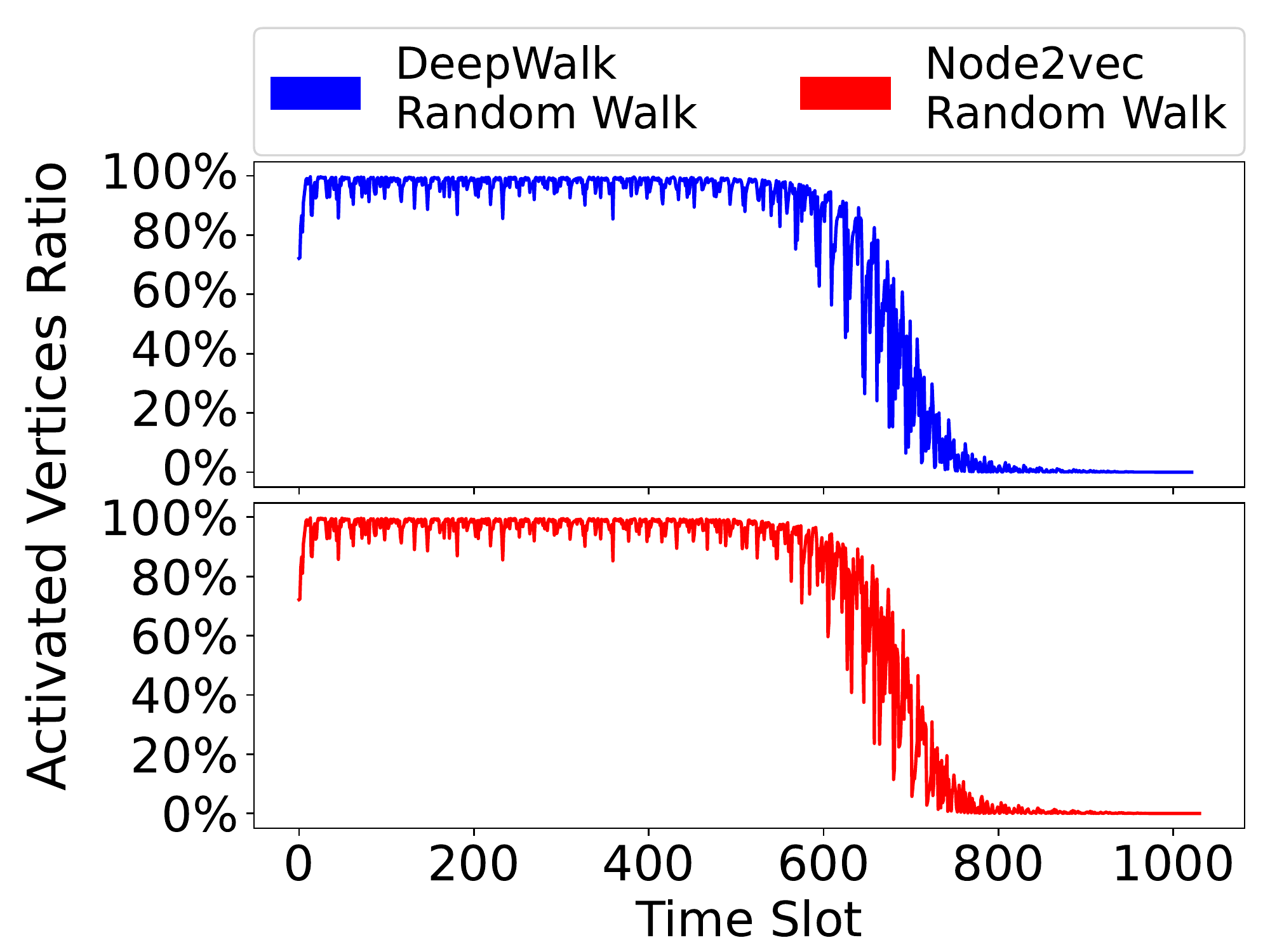}
    }
    \caption{The profile of running random walk tasks in SOGW system.}
    \label{fig:on-demand-load}
\end{figure}

However, none of the existing disk-based graph systems has considered the second-order random walk model to the best of our knowledge. In this paper, we aim to design a scalable disk-based graph system for executing second-order random walk models on large graphs. We also adopt the idea of processing the whole graph into blocks to address the insufficiency of memory. The main challenge of designing such a system is to deal with the extreme I/O overhead, which is twofold:

\textbf{Massive Light Vertex I/Os.} When generating a random walk in existing block-based graph systems, the walk is associated with the block which contains its current vertex. Since the block is loaded into memory before processing the walk, the current vertex and its neighbors are in the block, and it is efficient to update the first-order random walk without disk I/Os. However, when processing a second-order random walk which requires both the current vertex and previous vertex information, although the current vertex is easily retrieved from memory, the previous vertex might be retrieved from any other blocks in the disk, leading to a vertex I/O. These vertex I/Os are random and light and make it extremely I/O expensive to realize second-order random walk models on existing graph systems. Figure~\ref{fig:io-percentage} visualizes the costs of running DeepWalk (i.e., first-order) and Node2vec (i.e., second-order) random walk in SOGW system (introduced in Section~\ref{sec:expr_settings}) with three large graph datasets, and we decompose the cost into block I/Os, walk I/Os, vertex I/Os and walk updating costs. It is clear to see that, in the second-order random walk task, the efficiency bottleneck is the cost of vertex I/Os.

\textbf{Low Utilization of Block I/Os.} Most of the existing block-based graph systems load the whole block at once. However, when the workload of a random walk task is light, or when the task is about to finish, the activated vertex might be just a small portion of the whole block, leading to a waste of block I/Os. {Figure~\ref{fig:activated-ratio} visualizes the activated vertex ratio with regard to the time slots when running DeepWalk and Node2vec random walk tasks in SOGW system with LiveJournal dataset. We see that at the end of the tasks (i.e., about the last 20\% time slots), the ratio is close to zero.}   
To address low block I/O utilization, DynamicShards~\cite{Vora2016} and Graphene~\cite{Liu2017} dynamically adjust the layout of graph blocks to reduce the loading of useless data, but they do not consider the random walk features. GraphWalker~\cite{Wang2020} determines a proper block size according to the total number of random walks to improve the block I/O utilization, but such a solution is static and fails when tasks are about to finish and only few walks remain in the block.

To tackle the above two problems, we propose \sys, an \textit{I/O-efficient disk-based graph system for scalable second-order random walk}, which is equipped with a bi-block execution engine and a learning-based block loading model to improve the I/O efficiency. The main techniques are as follows.

\textbf{Bi-block Execution Engine.} To solve the problem of massive light vertex I/Os, we propose a bi-block execution engine, which keeps two blocks (i.e., current block and ancillary block) in memory, guaranteeing the previous and current vertices are in memory as well. To realize an I/O-efficient bi-block execution engine, we need to schedule the execution sequence of blocks and organize the walk states properly, to reduce the block I/Os as much as possible.
First, we theoretically analyze the hardness of block scheduling problems in the disk-based graph systems and discuss the influence between block scheduling strategies and the I/O cost. Then we introduce a \textit{triangular bi-block scheduling strategy} which eliminates half block I/Os compared to the standard scheduling strategy. Furthermore, we develop a \textit{bucket-based in-memory walk management} approach which merges random vertex I/Os into the sequence block I/Os, and a \textit{skewed walk storage} to ensure the correctness of the new scheduling strategy. 

\textbf{Learning-based Block Loading Model.} To improve the utilization of block I/Os, the challenge is to capture the dynamic workloads and estimate the costs of different block loading methods. In this work, we introduce two block loading methods -- full load and on-demand load. The former is the traditional block loading method, and the latter only loads activated vertices. Then we build a learning-based block loading model in \sys, which dynamically selects proper block loading methods based on online statistics. The model uses the linear regression method to learn cost estimation models for the two block loading methods from historical data and derives a simple threshold-based selection criterion.




Finally, combining with the above technical contributions, we carefully implement \sys~to efficiently process second-order random walk tasks on large graphs with a single machine. Experimental results on six large datasets show that \sys~ achieves efficiency improvement of more than one order of magnitude in common second-order random tasks such as random walk generation and PageRank query using Node2vec random walk model.
To summarize, our contributions are as follows:
\begin{itemize}
    \item We identify the I/O inefficiency of running second-order random walk models on existing disk-based graph processing systems and propose an I/O-efficient system \sys.
    \item We propose an efficient bi-block execution engine, which equips a triangular bi-block scheduling strategy, skewed walk storage, and bucket-based in-memory walk management to eliminate massive vertex I/Os.
    \item We propose a learning-based block loading model to improve the block I/O utilization when a few walks remain in the bucket.
    \item {We compare our \sys~with SOGW and \cache~on real-world and synthetic large graphs. The results show that \sys~significantly reduces the end-to-end time of second-order random walk tasks and improves the I/O efficiency. We also demonstrate the effectiveness of \sys~for the first-order random walk.}
\end{itemize}

\section{Preliminary}








A graph $G=(V,E)$ is defined by a set of vertices $V$, and a set of edges $E$. Each edge is a pair of the form $e = (u, v), u, v \in V$, where $u$ is the source vertex and $v$ is the destination vertex of $e$, and $a_{vz}$ represents the corresponding weight. If such $e = (u,v)$ exists, then $v$ is a neighbor of $u$, and we use $N(u)$ to denote the set of neighbors of $u$. In disk-based graph systems, a graph is partitioned into several blocks, and we use $B(v)$ to denote the ID of the block which vertex $v$ belongs to, and the $i$th block is denoted as $B_i$. Given a graph partition, we use $N_B$ to represent the number of partitioned blocks. 
The notations frequently used in this paper are listed in Table \ref{tab:symbols}.

\begin{table}[t]
\centering
\scalebox{0.85}{
\begin{tabular}{l|l}
\hline
\textbf{Symbol} & \textbf{Description}                                               \\ 
\hline \hline
$G=(V,E)$    & Graph $G$ with a set of vertices $V$ and a set of edges $E$. \\ \hline
$e = (u,v)$     & An edge from $u$ to $v$.                                        \\ \hline
$a_{uv}$        & The weight of the edge $(u,v)$.                            \\ \hline
$N(u)$          & The set of neighbors of $u$.                                   \\ \hline
$B(v)$          & The ID of the block which vertex $v$ belongs to.                \\ \hline
$B_i$           & The block whose ID is $i$ .                               \\ \hline
$b_i$           & The bucket whose ID is $i$ .                               \\ \hline
$N_B$           & The total number of partitioned blocks of graph $G$.            \\ \hline
$w$, $\mathbb{W}$ & A walk and a set of walks, i.e., $\mathbb{W}=\{w\}$.            \\ \hline
$w^v$           & A walk which currently resides on vertex $v$.                   \\ \hline
$w_u^v$         & \tabincell{l}{A walk whose current vertex is $v$, and previous vertex is $u$.} \\ \hline
$\mathbb{A}$    & The set of activated vertices.                                 \\ \hline
$t_f$           & \tabincell{l}{Total time of block loading and executing stage with the \\full-load method.} \\ \hline
$t_o$           & \tabincell{l}{Total time of block loading and executing stage with the \\on-demand load method.} \\ \hline
$\eta_0$        & The threshold of selecting block loading method.                               \\ \hline
\end{tabular}}
\caption{The symbols frequently used in this paper.}
\label{tab:symbols}
\end{table}

\subsection{Random Walk}
A random walk $w$ on graph $G=(V, E)$ starts from a vertex, and for each step, it selects the next vertex to visit following a transition probability distribution $p$. 
In first-order random walk models, $p=p(z|v)$, which means the selection of the next vertex $z$ only depends on the vertex $v$ that walk $w$ currently resides on. A walk $w$ currently residing on vertex $v$ is denoted by $w^{v}$, and $v$ is called the current vertex of $w$. 
In second-order random walk models, $p=p(z|uv)$, where $u$ is the vertex that walk $w$ previously resided on, and $v$ is the current vertex of $w$. Such a walk is denoted by $w_{u}^{v}$ and the corresponding distribution is called edge-edge distribution. Following the edge-edge distribution, selecting the next vertex $z$ depends on both vertex $u$ and $v$.

Next we briefly review two popular random walk models.

\textbf{DeepWalk model}. In this paper, the DeepWalk model represents the first-order random walk model used by DeepWalk, a method of learning graph embeddings. 
The transition distribution in DeepWalk model is $p(z|v)=a_{vz}/Z_v$, where $Z_v = \sum_{t \in N(v)}a_{vt}$. The same distribution is used in most other first-order random walk models.

\textbf{Node2vec model}. In this paper, the Node2vec model represents the second-order random walk used by Node2vec, which is also a method of learning graph embeddings.
In this model, for walk $w_u^{v}$, we define biased weight:
\begin{equation}
    a_{vz}^{\prime} = \left\{
    \begin{array}{rcl}
    \frac{a_{vz}}{p} & h_{uz} = 0 \\
    a_{vz} & h_{uz} = 1 \\
    \frac{a_{vz}}{q} & h_{uz} = 2
    \end{array}\right.
    \label{equ:node2vec}
\end{equation}
where $z \in N(v)$, $p, q \in \mathbb{R}^{+}$ are two hyperparameters, and $h_{vz}$ is the shortest hops between $v$ and $z$. For edges $(v, u)$ and $(u,z)$, the edge-edge transition distribution $p(z|uv) = a_{vz}^{\prime} / Z_{v}^{\prime}$, where $Z_{v}^{\prime} = \sum_{t \in N(v)}a_{vt}^{\prime}$.


\begin{figure}[t]
    \centering
        \scalebox{0.35}{
        \includegraphics{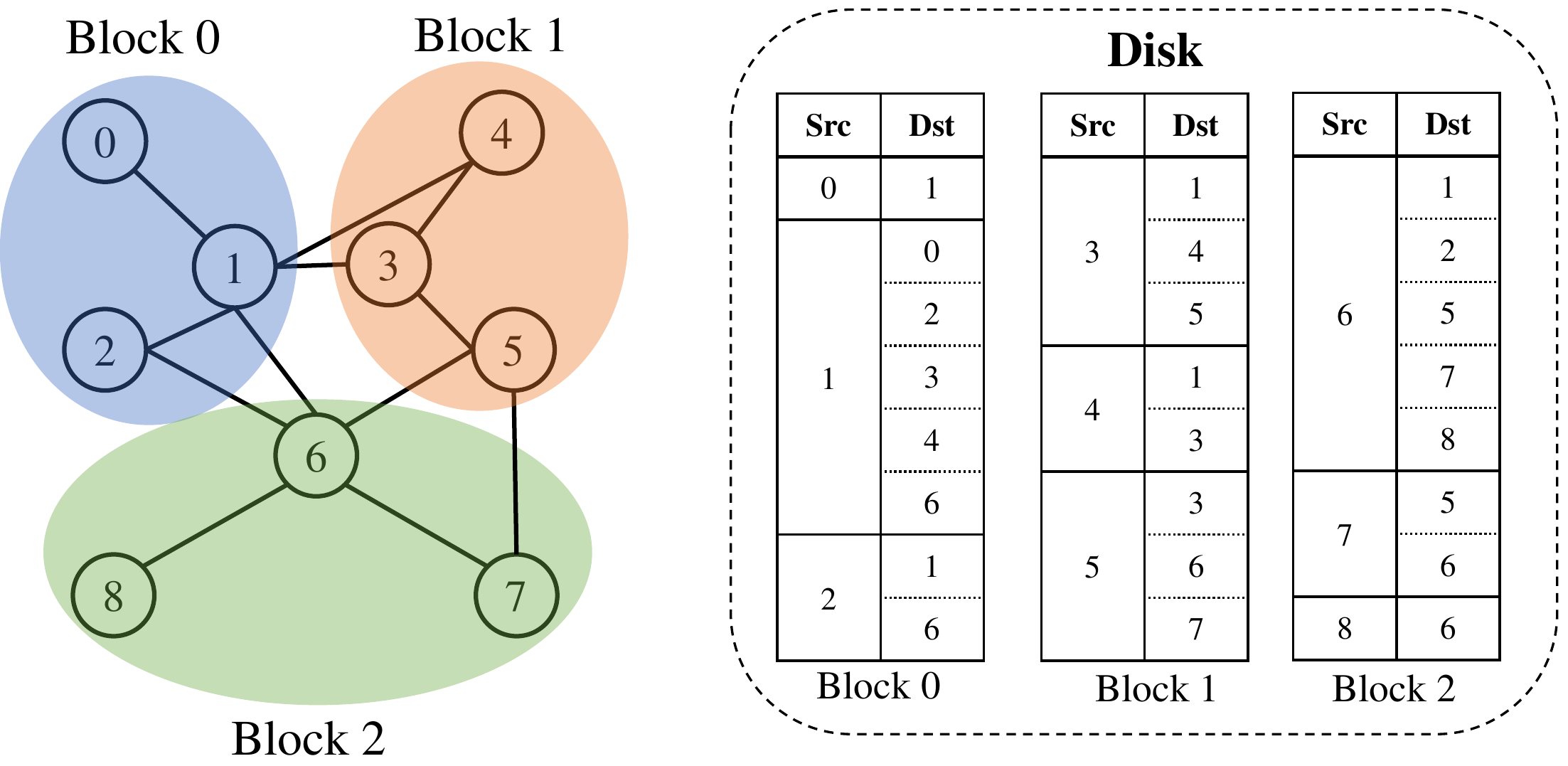}
    }
    \caption{A partitioned graph and its storage in disk.}
    \label{fig:example}
\end{figure}
\subsection{Disk-based Random Walk System} \label{sec:disk-based-rw-systems}
In disk-based random walk systems, a graph is partitioned into several blocks, and only one block is loaded into memory at a time while updating the corresponding random walks. {Figure~\ref{fig:example} illustrates a partitioned graph and its storage in disk.}
GraphWalker~\cite{Wang2020} is a state-of-the-art disk-based random walk system. It first loads a block into memory, then loads walks belonging to that block and updates them asynchronously. 
These walks are called the \textit{current walks}, and the block loaded into memory is called the \textit{current block}. It applies a state-aware block scheduling strategy, which chooses the block with most walks in it to be the current block. The updating of a walk stops either when it moves out of the current block or when it reaches the termination condition. For the former situation, the walk is associated with the new block where it currently moves into. After updating all walks in the current block, the system chooses the next current block and updates the corresponding walks in memory. Such a cycle is called a \textit{time slot}. 

As introduced in the Introduction, there will be extreme I/O overhead when realizing second-order random walk models on existing disk-based random walk systems because of the massive light vertex I/Os and low utilization of block I/Os. {In the next section, we will introduce \sys, which is an I/O-efficient disk-based graph system for the scalable second-order random walk over large graphs.}
\section{Overview of \sys}
\label{sec:overview}
\begin{figure}[t]
    \centering
    \includegraphics[width=8.5cm]{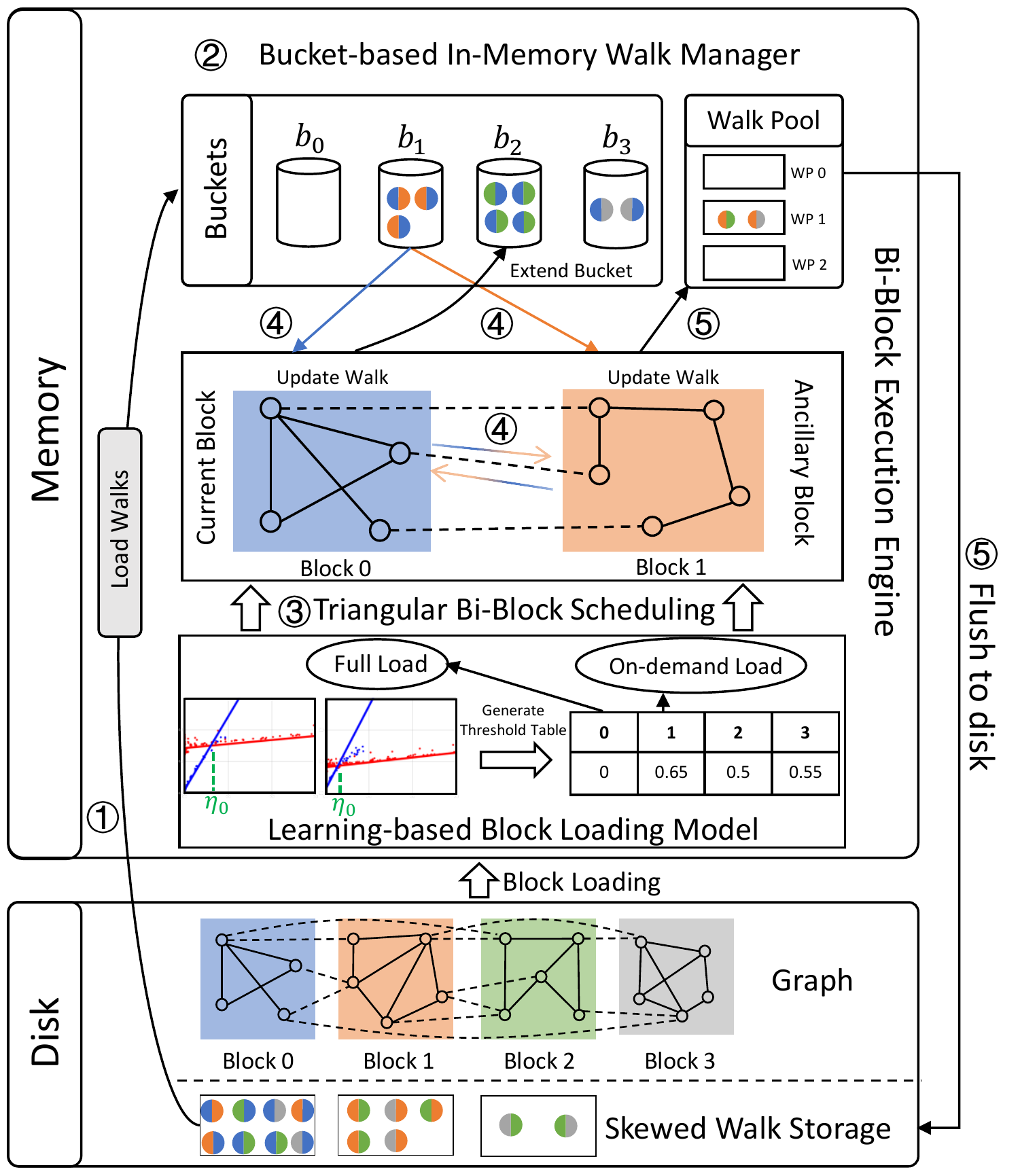}
    \caption{The execution flow of \sys. The semi-circle represents a vertex in a walk, and the left semi-circle is the previous vertex while the right one is the current vertex. The color of a semi-circle indicates the block that the vertex belongs to. For example, the orange semi-circle means the vertex is from block 1. Best viewed in color.}
    \label{fig:overview}
\end{figure}

\sys ~is an I/O-efficient disk-based graph system for the scalable second-order random walk. 
Similar to previous works, the graph and intermediate walks are stored on disk. The graph is partitioned into blocks, and each block is associated with a walk pool storing the intermediate walks. 
The difference is as follows: to reduce the massive vertex and block I/Os, we design a \textit{bi-block execution engine} and a \textit{learning-based block loading model}. Figure \ref{fig:overview} describes the high-level execution flow of \sys. During the execution, the bi-block execution engine iteratively selects a block as the current block, uses the learning-based block loading model to load an ancillary block into memory, and updates the intermediate walks associated with the current block. Next, we detailedly present the execution flow of \sys~ in a time slot.



In each time slot, \textcircled{1} the engine uses the \textit{bucket-based in-memory walk manager} to load the intermediate walks associated with the current block into memory and merges them with the one in the in-memory walk pool forming the \textit{current walks}. \textcircled{2} Then the manager splits current walks into buckets, and each bucket stores the walks having the same block set, in which the pair of blocks contains their previous and current residing vertices. Such bucket-based in-memory walk management merges massive vertex I/Os into a single block I/O, and the details are described in Section~\ref{subsec:walk_managment}.   

Considering that the previous and current residing vertices are involved in two blocks, before processing a bucket, we need to load another block into memory, called the ancillary block. In \sys, each ancillary block is corresponded with a bucket, while the current block is shared among all buckets. \textcircled{3} In each time slot, the bi-block execution engine uses \textit{triangular bi-block scheduling} method to determine the loading sequence of ancillary blocks and also uses the learning-based block loading model to load the block. 


After loading the blocks, \textcircled{4} the engine asynchronously updates the walks in the bucket. The current vertex of walks in this bucket can be either in the current block or the ancillary block, as the blue and orange arrows show. Moreover, since there are some edges connecting the two blocks, walks can also be updated across the two blocks. The updating of a walk stops when it moves to any vertex not belonging to the blocks in memory or when the termination condition is reached. \textcircled{5} For the former situation, walk persistence is needed to preserve the information of these intermediate walks and update them in future. Intermediate walks have two places to go. Most of them are stored in the in-memory walk pool with skewed walk storage (introduced in Section~\ref{sec:walk-storage}), and others may be moved into buckets, caused by bucket-extending strategy introduced in Section \ref{sec:walk-processing}. When the size of a walk pool reaches the pre-defined threshold, the in-memory walk pool is flushed to disk. 
After all walks in the bucket have terminated or been persisted, the next ancillary block is chosen with triangular bi-block scheduling, and the corresponding bucket is executed iteratively.



{Note that the learning-based block loading model in \sys~is proposed to improve the I/O utilization when the number of current walks is small. It uses a linear regression model to predict the cost by learning from historical logs, and on top of the model, we derive thresholds for each block and use the threshold to select the block loading method}. The details are introduced in Section~\ref{sec:io-demand}.
In Figure \ref{fig:overview}, block 0 is fully loaded, and block 1 is loaded with the on-demand load method. 

\section{Bi-Block Execution Engine}


The basic idea of the bi-block execution engine is to keep two blocks (i.e., current block and ancillary block) in memory, thus guaranteeing the current and previous vertices are both in memory. To establish such an engine, we need to address the following two research questions: (RQ1) how do we schedule the two blocks to reduce block I/Os? and (RQ2) how do we manage the states of walks in disk and memory to support I/O-efficient block scheduling?

\subsection{Block I/O Cost Analysis and Scheduling Strategy Revisit}
\label{subsec:revisit}
Given a second-order random walk task and its input, the total number of block I/Os in \sys~is related to two factors: the number of current block I/Os (i.e., time slots) and the number of ancillary block I/Os in each time slot. Minimizing the total number of block I/Os can be achieved by reducing the current block I/Os and the ancillary block I/Os separately.
However, we show that obtaining the minimal number of current block I/Os is an NP-hard problem. In the following subsection, we present our triangular bi-block scheduling strategy, which focuses on reducing ancillary block I/Os.

Different block scheduling strategies incur different numbers of current block I/Os.
We define the optimization problem of minimal current block I/Os as below:  
\begin{myDef} \textbf{(The minimal current block I/Os problem).} 
Given a graph $G=(V,E)$ which is partitioned into $N_B$ blocks $\mathbb{B}=\{B_1, B_2, ..., B_{N_B}\}$, and a task of the second-order random walk which generates $M$ random walks, where each random walk $w_i$ has a certain sequence of accessing current blocks, denoted by $\{B_{i0}, B_{i1}, ..., B_{ij}\}$, $0<i\le M$, $B_{ij}\in \mathbb{B}$. The goal is to find a global block scheduling strategy $\{B_1, B_2, ...., B_K\}$ so that all $M$ random walks are generated and $K$ is minimized, where $K$ equals the number of current block I/Os.
\end{myDef}

The following theorem presents the hardness of the problem.
\begin{myTheorem} 
The minimal current block I/Os problem is NP-hard.
\label{th:hardness}
\end{myTheorem}

\begin{myProof*}
{
Given a sequence $S = s_1, s_2, ..., s_m$, a \textit{supersequence} $S^\prime$ is a sequence from which items (e.g., $s_i$) can be removed to produce $S$. We write $S^\prime > S$ if $S^\prime$ is a supersequence of $S$. Given a set $R=\{S_1, S_2, ..., S_M\}$ of sequences, the \textit{Shortest Common Supersequence} of $R$, SCS($R$), is a shortest sequence $S^\prime$ such that $S^\prime>S_i, i = 1, 2, ..., M$.}

{In the minimal current block I/Os problem, the current block access sequence associated with a random walk $w$, $\{B_{0}, B_{1}, ..., B_{m}\}$, forms a sequence $S = s_1, s_2, ..., s_m$, {where $s_i = B_i$, and we use $S \sim w$ to represent the above relationship between the random walk $w$ and sequence $S$.}}

{For a set $\mathbb{W}=\{w_1, w_2, ..., w_M\}$ of $M$ random walks, their sequences of current blocks form a set $Q = \{S_1, S_2, ... S_M\}$, {where $S_i \sim w_i$}. By adding the constraint that $R = Q$, the SCS problem is reduced to the minimal current block I/Os problem. Previous studies~\cite{scs_theory_1,scs_theory_2} show that SCS is an NP-Complete problem, therefore, the minimal block I/Os problem is an NP-hard problem.
}
\end{myProof*}

Although there are several approximation algorithms for the SCS problem~\cite{TURNER19891}, they assume the sequences are known ahead. However, in our problem, the block access sequence for a second-order random walk is unknown, and we need to design an online algorithm to solve the above problem. As far as we know, most existing heuristic online solutions of SCS problem have no {(or poor)} approximation error bound of the optimal solutions.
{Therefore, we empirically studied the different scheduling strategies~\cite{4673529,Wang2020} for current blocks; the results in Appendix~\ref{app:scheduling} show that no single method performs optimally on all datasets, and the performance of the same method on different datasets may vary widely.
But in general, the Iteration-based method, which loads the block from $B_0$ to $B_N$ iteratively, achieves the best result in most cases.} With such observations, in this paper, {we adopt the Iteration-based method to schedule the current block,} and then focus on developing a new scheduling strategy, which optimizes the ancillary block I/Os.

\subsection{Triangular Bi-Block Scheduling based Execution (RQ1)} 
\label{sec:block-scheduling}
{As mentioned before, we use the Iteration-based method to schedule the current blocks, which sequentially loads the current blocks from $B_0$ to $B_{N_B - 1}$ iteratively into memory, and skips the loading of a current block if there is no intermediate walk in it. }
Due to the asynchronous walk updating method~\cite{Wang2020} in \sys, there is no walk whose previous vertex and current vertex are in the same block. Then for each current block, we at most process $N_B - 1$ ancillary blocks, which incurs $N_B - 1$ block I/Os in a time slot. In other words, the total block I/Os of processing the whole graph once is at most
\begin{equation}
    N = N_B + N_B(N_B - 1)
      = N_B^2.
\label{eq:standard}
\end{equation}

With the help of our skewed walk storage introduced in the next subsection, we can only load ancillary blocks whose ID is larger than the one of the current block. This is the new triangular bi-block scheduling strategy, and the total block I/Os is computed as follows: 
 \begin{equation}
    N = N_B - 1 + \sum_{b=0}^{N_B-2}(N_B - 1 - b)
      = \frac{1}{2}(N_B + 2)(N_B - 1).
      \label{eq:bi-block}
\end{equation}
Compared to the Equation~\ref{eq:standard}, the triangular bi-block scheduling strategy saves about 50\% block I/Os.

Algorithm \ref{alg:triangular-scheduling} illustrates the execution procedure on the basis of the triangular bi-block scheduling strategy. The current block ID $b$ iterates from $0$ to $N_B - 2$ (Line \ref{line:current-block}), and in each time slot the ancillary block ID iterates from $b + 1$ to $N_B - 1$  (Line \ref{line:ancillary-block}). After choosing the current block, the associated walks are loaded into memory and collected into different buckets (Line \ref{line:load-walks}). The details of bucket collection is described in Section \ref{sec:in-memory-walk-management}. Finally, walks are processed in bucket id order (Line \ref{line:processwalk}), and the update of walks in each bucket can be accelerated in parallel. {We discuss the space and time complexity of Algorithm \ref{alg:triangular-scheduling} in Appendix \ref{app:complexity}.
Note that the correctness of Algorithm~\ref{alg:triangular-scheduling} is guaranteed by our skewed walk storage, which is introduced in Section~\ref{sec:walk-storage}. And the detailed proof of the correctness is presented in Appendix \ref{app:correctness}}.

\begin{algorithm}[t]
\caption{Triangular Bi-Block Scheduling in \sys}
\begin{algorithmic}[1]
\While{has unfinished walk}
    \For{$b = 0 \to N_B - 2$} \label{line:current-block}
        \State{$curWalks[] \gets$ \Call{LoadWalks}{$b$}} \label{line:load-walks} \algorithmiccomment{From the skewed walk storage}
        \For{$w \in curWalks[]$}\algorithmiccomment{Collect bucket}
            \If{\Call{PreBlockId}{$w$} $= b$}
                \State{$p \gets$ \Call{CurBlockId}{$w$}}
            \Else
                \State{$p \gets$ \Call{PreBlockId}{$w$}}
            \EndIf
            \State{$bucket[p] \gets bucket[p] \cup w$}
        \EndFor
        \State\Call{LoadSubGraph}{$b$} \label{line:load-current-block}
        \For{$i = b + 1 \to N_B - 1$} \label{line:ancillary-block}
            \State\Call{LoadSubGraph}{i} \label{line:load-ancillary-block}
            \For{$w \in bucket[i]$}
                \State\Call{ProcessWalk}{$w, b, i, bucket$} \algorithmiccomment{Algorithm~\ref{alg:walk-processing}} \label{line:processwalk}
            \EndFor
        \EndFor
    \EndFor
\EndWhile
\end{algorithmic}
\label{alg:triangular-scheduling}
\end{algorithm}

\subsection{Walk Management and Processing (RQ2)}
\label{subsec:walk_managment}
In this subsection, we first describe the skewed walk storage, which supports the triangular bi-block scheduling strategy, and then introduce the bucket-based in-memory walk storage, which helps cluster the random vertex I/Os into blocks.

\subsubsection{Skewed Walk Storage} \label{sec:walk-storage}
{Traditional walk storage methods associate a walk with the block to which its current vertex belongs. This brings limitations when updating walks under the triangular bi-block scheduling strategy. 
First, suppose that $B_b$ is the current block and $B_p$ is the previous block. With the traditional walk storage method, only walks currently in block $B_b$ are loaded into memory, so only walks $w_u^v$ such that $u \in B_p$, $v \in B_b$ get updated in the triangular bi-block scheduling strategy. The walks $w_u^v$ such that $u \in B_b$, $v \in B_p$ are still in disk, and cannot utilize the ancillary block which has been loaded into memory more efficiently.
Second, traditional walk storage cannot correctly support the triangular bi-block scheduling strategy. Because the walks currently in block $B_b$ might have the ones of which the block ID of previous vertex 
is smaller than the ID of $b_b$, then these walks would never be updated in triangular bi-block scheduling strategy.
Therefore, we design a simple but effective skewed walk storage, which not only supports the triangular bi-block scheduling strategy but also helps update as many walks a possible in a time slot.}

The skewed walk storage in \sys~ takes both the previous and current vertex of the walk into consideration to arrange the walks.
Specifically, a walk $w_u^v$ is associated with block $B_i$, where $i = min\{B(u), B(v)\}$. 
Compared to the traditional walk storage, such storage splits the walks whose current vertices belong to the same block into two groups. One group contains the walks whose current vertices belong to the block with larger ID than their previous vertices belong to; the other group contains the remaining walks. Consequently, in the context of the triangular bi-block scheduling strategy, the first group is processed when the block which current vertices belong to is loaded as an ancillary block, and the second group is processed when the block which current vertices belong to is loaded as the current block.




\subsubsection{Bucket-based in-Memory Walk Management} \label{sec:in-memory-walk-management}
As introduced in Section~\ref{sec:overview}, to merge random vertex I/Os into block I/Os, the bucket-based in-memory walk manager splits the current walks into buckets, and each bucket stores the walks having the same block set in which the pair of blocks contain their previous and current residing vertices. Specifically,
let $\mathbb{b} = \{b_i, 0 \leq i < N_B\}$ be the set of buckets, then with the skewed walk storage, the current walks might also contain walks whose previous vertex belongs to the current block of the time slot. Let $B_i$ be the current block, then a walk $w_u^v$ is distributed into bucket:
\begin{equation} \label{equ:walk-management}
    \left\{
    \begin{array}{rcl}
    b_{B(v)} & \text{if} & u \text{~belongs to block~} B_i\\
    b_{B(u)} & \text{if} & v \text{~belongs to block~} B_i
    \end{array}\right.
\end{equation}
That is to say, the bucket collection also relies on both the current vertex and the previous vertex of the walk. Furthermore, combined with the skewed walk storage, if the walk is collected to bucket according to its previous vertex, then the ID of the block to which its current vertex belongs is smaller than that of the previous vertex, and vice versa. This walk management supports the triangular bi-block scheduling strategy.

\subsubsection{Walk Processing}
\label{sec:walk-processing}
\begin{algorithm}[t]
\caption{Walk processing in \sys}
\textbf{Parameters:} walk: $w$, current block ID: $b$, ancillary block ID: $i$
\begin{algorithmic}[1]
\Function{ProcessWalk}{$w, b, i, bucket[]$} \label{line:process-walk}
    \State{$w^{\prime} \gets \Call{UpdateWalk}{w, b, i}$}
    \State{$cur \gets \Call{CurBlockId}{w^{\prime}}$}
    \If{$cur < b$} \label{line:ass-con-1}
        \State{\Call{AssociateWithBlock}{$w^{\prime}, cur$}} \label{line:associate-walk-1}
    \ElsIf{$b < cur < i$} \label{line:ass-con-2-3}
        \If{\Call{PreBlockId}{$w^{\prime}$} = b}
            \State\Call{AssociateWithBlock}{$w^{\prime}, b$} \label{line:associate-walk-2}
        \Else
            \State\Call{AssociateWithBlock}{$w^{\prime}, cur$} \label{line:associate-walk-3}
        \EndIf 
    \ElsIf{$cur > i$} \label{line:ass-con-last-start}
        \If{\Call{PreBlockId}{$w^{\prime}$} = b}
            \State{$bucket[cur] \gets bucket[cur] \cup w^{\prime}$}\label{line:move2bucket}\algorithmiccomment{{\footnotesize Bucket-Extending}}
        \Else
            \State\Call{AssociateWithBlock}{$w^{\prime}, i$} \label{line:associate-walk-4}
        \EndIf
    \EndIf \label{line:ass-con-last-end}
\EndFunction
\State
\Function{UpdateWalk}{$w, b, i$} \label{line:update-walk}
    \While{\Call{CurBlockId}{$w$} = $b$ or $i$ and walk not terminated} \label{line:update-walk}
        \State{$w \gets \Call{SampleDestVertex}{w}$}
    \EndWhile
    \State\Return{$w$}
\EndFunction
\end{algorithmic}
\label{alg:walk-processing}
\end{algorithm}
Finally, we describe the procedure of walk processing by combining the techniques of triangular bi-block scheduling and bucket-based walk management in Algorithm \ref{alg:walk-processing}.
The association between updated walks and blocks follows the organization in the skewed walk storage, denoted by the function \textsc{AssociateWithBlock}, in which the walks are stored in the walk pool corresponding to the given block.
In Function \textsc{ProcessWalk} in Algorithm \ref{alg:walk-processing}, we first update the old walk $w$, and the new walk after updating is denoted as $w^{\prime}$. Here we use $pre$ and $cur$ to denote the previous block ID and the current block ID of the new walk $w^{\prime}$, and $b$ and $i$ to represent the current block ID and ancillary block ID which are in memory now.
For the new walk $w^{\prime}$, if $cur < b$ then it should be associated with block $cur$, since $(pre = b~\text{or}~i) > cur$, as shown in Line \ref{line:ass-con-1} and Line \ref{line:associate-walk-1}.
If $cur$ is between $b$ and $i$, the association depends on whether its previous vertex belongs to the current block or the ancillary block, as shown in Line \ref{line:ass-con-2-3} to \ref{line:associate-walk-3}. 
{As Line \ref{line:ass-con-last-start} to \ref{line:ass-con-last-end} shows, when the new walk moves to the block whose ID is larger than the one of the ancillary blocks in memory (i.e., $i$), we associate it with the block where the new walk previously moves out from. An exception is shown in Line \ref{line:move2bucket}, where we develop a \textit{bucket-extending} strategy to improve the efficiency further. Specifically, when the new walk is previously moved out from the current block $b$ (i.e., $pre$=$b$), instead of associating the new walk $w^{\prime}$ with the current block $b$, we add it to the bucket corresponding to the current block $cur$ of the new walk, which has not been executed as an ancillary block yet in this time slot.}
The bucket-extending strategy ensures new walks who meet the above condition are able to update as many steps as possible in a time slot. However, it also brings synchronization overhead when the updating of walks is executed in parallel. The efficient implementation of bucket-extending is discussed in Section \ref{sec:parallelism}.

During the updating, since there are two blocks in memory, the walks keep moving when they jump between the two blocks in memory, as Line \ref{line:update-walk} in Function \textsc{UpdateWalk} shown. Therefore, if two blocks are strongly connected (i.e., they has many edges across them), then walks can update much faster, without swapping the blocks in memory.

\begin{figure}[t]
    \centering
    \includegraphics[width=0.5\textwidth]{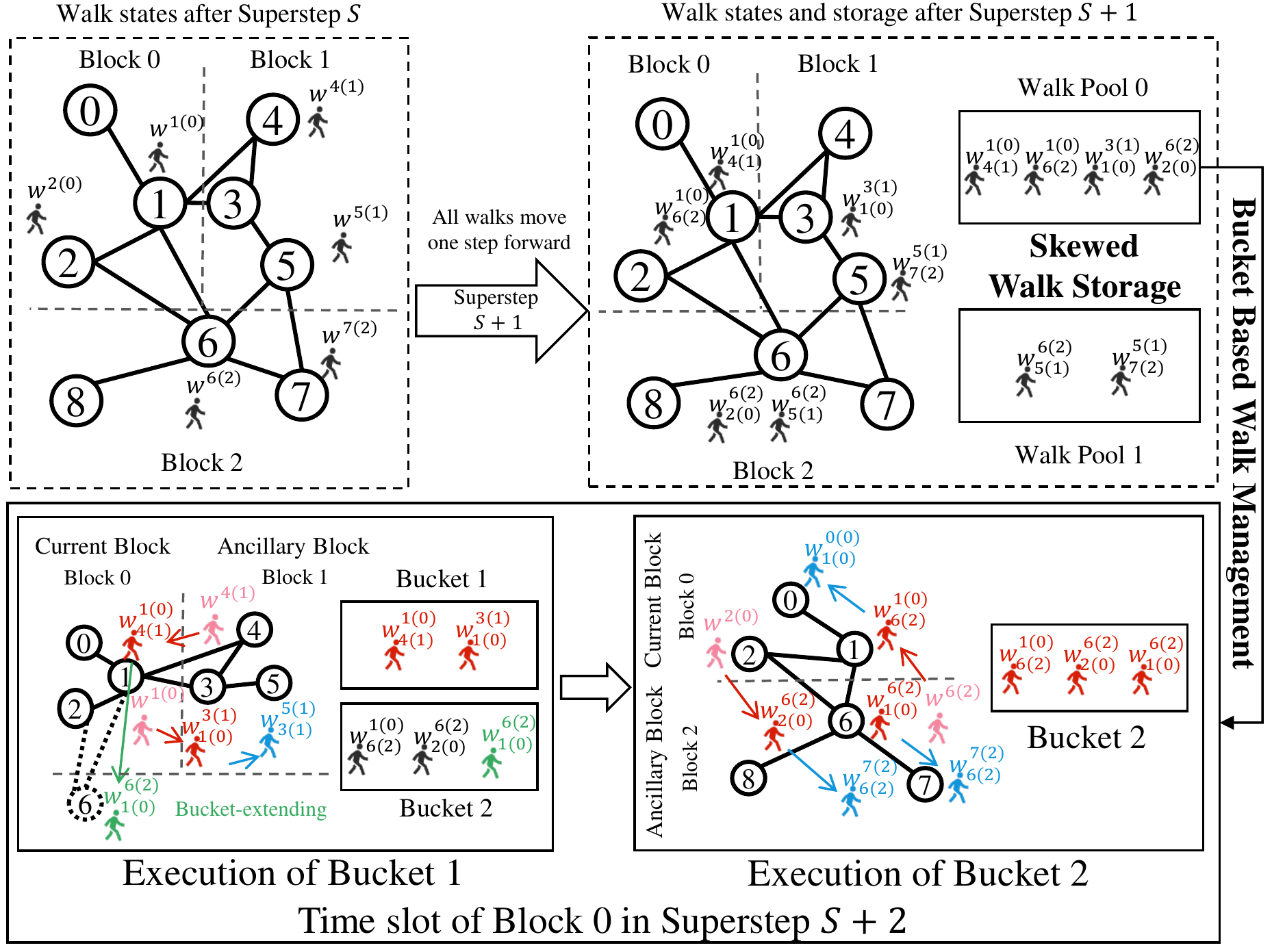}
    \caption{An example of walk processing with the bucket-based in-memory walk management.}
    \label{fig:engine}
\end{figure}

{Figure \ref{fig:engine} illustrates the key procedures of the walk processing with the skewed walk storage, bucket-based walk management, and the execution of buckets. Here we use $w_{u(pre)}^{v(cur)}$ to denote a walk, where $pre = B(u)$ and $cur = B(v)$. A Superstep shown in the figure represents the procedure that all walks in the task moves at least one step forward. In Superstep $S + 2$, the walkers with red color are the ones being updated in their corresponding time slots, and the walkers in pink represent where the red walkers come from. The blue walkers represent where the red walkers are going to visit in the next step, and these walkers can be updated further in their time slots. The green walker is similar to the blue ones, but they have moved out the blocks in memory. In this example, the green walker satisfies the condition of the bucket extending strategy.} 

\section{Learning-based Block Loading Model}
\label{sec:io-demand}
{As discussed in Section~\ref{sec:block-scheduling}, the majority of block I/Os are caused by the ancillary block loading. When only a small portion of vertices in a block have walks residing on, then it may lead to a waste of the block I/Os. To improve the I/O-utilization, we introduce two block loading methods in \sys, namely full load and on-demand load, and propose a learning-based model to automatically select a block loading method {for ancillary block} according to the runtime statistics.}

\subsection{Full load and On-Demand Load}
\hspace{2ex}\textbf{Full-Load Method.} This method has been widely used in existing disk-based graph systems, and it means that a whole block is loaded into memory at once. In \sys, the slice of \textit{Index File} and \textit{CSR File} of the corresponding block is loaded into memory. For example, if block 1 of the partitioned graph shown in Figure \ref{fig:example} is loaded by the full-load method, then the grey cells in Figure \ref{fig:CSR} are loaded into memory. 
\textbf{On-Demand-Load Method.} This method means that only activated vertices in the corresponding block are loaded into memory. To load a block $B$ with the on-demand-load method, we first check the current vertex and the previous vertex of each walk in the walk set $\mathbb{W}$, and record all the vertices which belong to the block $B$. These vertices are the activated vertices and will be used to update the walks. 
For each ancillary block, $\mathbb{W}$ is composed of the walks in the corresponding bucket. 
Then only the CSR segmentation related to the activated vertices is loaded.
In \sys, the on-demand loading happens right before the execution of each bucket (as a replacement of Line~\ref{line:load-ancillary-block} in Algorithm~\ref{alg:triangular-scheduling}). Note that if any walk is able to update more than one step during execution and the information of the current vertex is not in memory, we should get its CSR segmentation solely from disk, which incurs few random vertex I/Os, and store it in memory as well. 

\begin{figure}
    \centering
    \subfigure[Full-load of block 1.]{
        \label{subfig:on-demand-load-block1}
        \includegraphics[width=0.22\textwidth]{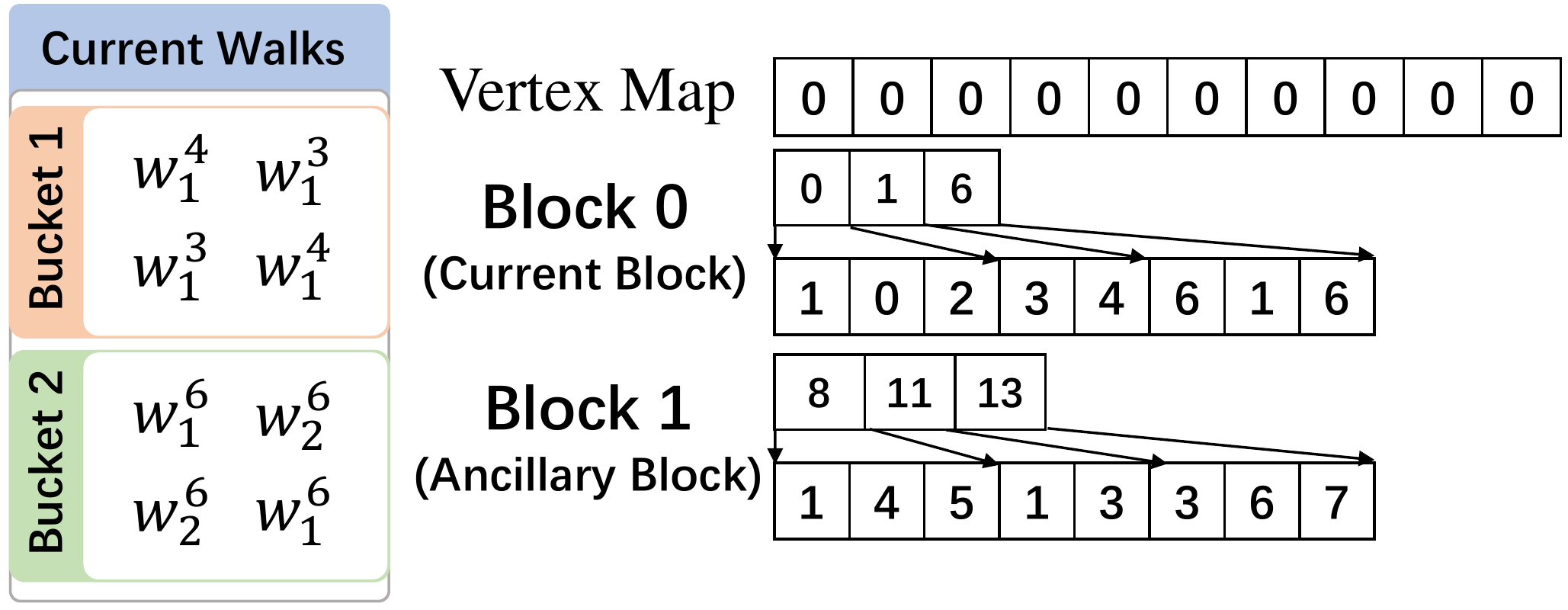}
    }
    \subfigure[On-demand-load of block 2.]{
        \label{subfig:on-demand-load-block2}
        \includegraphics[width=0.22\textwidth]{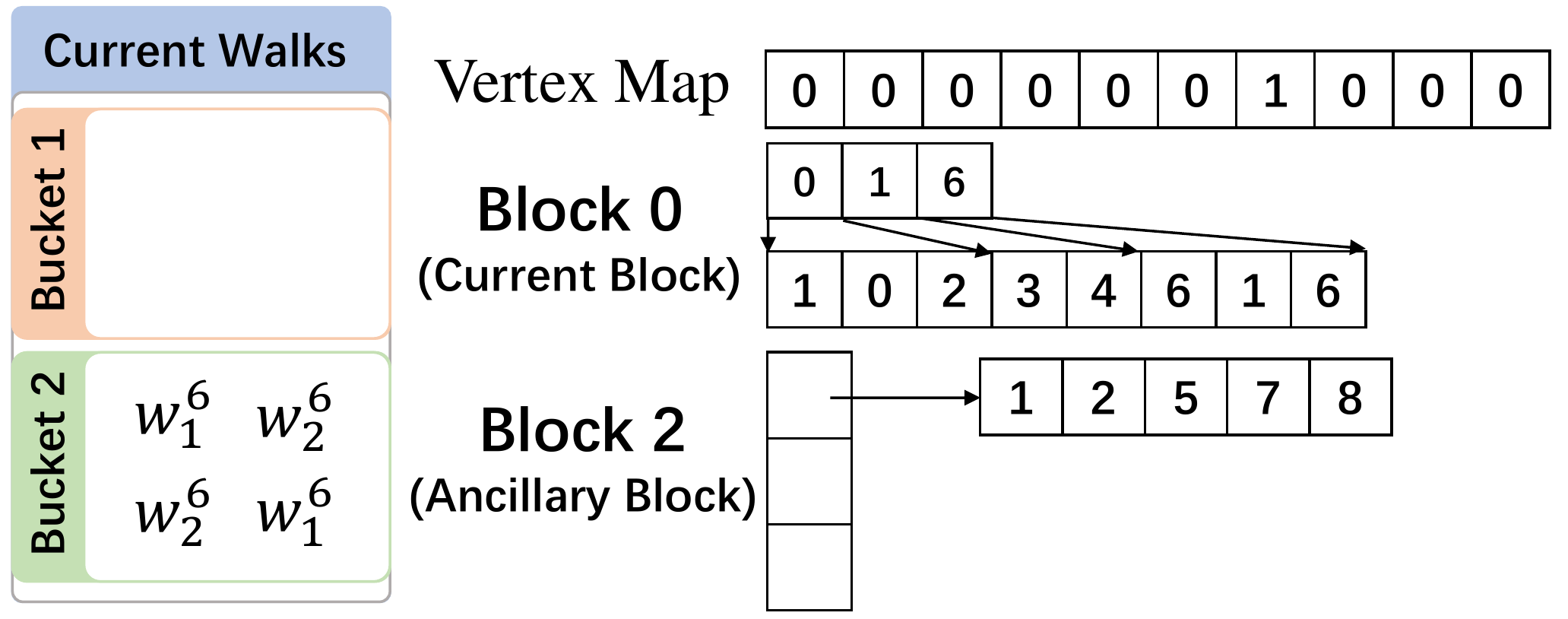}
    }
    \caption{An example of different block loading methods. \textit{Vertex Map} is used to record the activated vertices. In this example, block 0 is the current block. During execution, block 2 is loaded with the on-demand-load method, and block 1 is loaded with the full-load method.}
    \label{fig:on-demand-load}
\end{figure}

\textbf{Example.} In Figure \ref{fig:on-demand-load}, we compare the I/O difference between full load and on-demand load through an example. {Assume that there are eight walks being the current walks, and each value stored in the Index File and CSR File occupies 4 bytes in disk. The system decides to load block 2 with the on-demand-load method and load block 1 with the full-load method. 
The \textit{Vertex Map} is used to record the activated vertices. 
Since block 1 is decided to be loaded with the full-load method as shown in Figure \ref{subfig:on-demand-load-block1}, the whole slice of the index file and the CSR file are loaded into memory, incurring 32 bytes I/O. After executing updates of walks in bucket 1, the memory for block 1 is freed. Before executing bucket 2, all walks in it are scanned to tally activated vertices for block 2. In the example shown in Figure \ref{subfig:on-demand-load-block2}, only the information of vertex 6 is needed, so the system only loads the CSR segmentation of vertex 6 into memory, and 20 bytes I/O is needed. In total, 52 bytes I/O is required to load the CSR information for the ancillary blocks. However, 64 bytes disk I/Os would be incurred if the pure full-load method is used to load both block 1 and block 2. In this example, block I/Os are saved by 18.8\% by mixing the full load and on-demand load methods. Furthermore, there is no need to allocate memory to store the slice of the index file of block 2. This example implies that it is worthy of making a trade-off between two block loading methods.}

\subsection{Linear Regression Model for Block Loading}
The key to automatically selecting a loading method for a block is to estimate the corresponding cost. However, it is difficult to develop heuristics for the cost model since the number of activated vertices is task-dependent and different data structures between the two block loading methods also influence the efficiency. In this paper, we develop a learning-based model to predicate the cost of each loading method. Next, we will describe the model and its training method.
\subsubsection{Linear Regression Models for Cost Estimation}
A block processing can be divided into two stages, the loading stage and the executing stage. 
Under full-load mode, to process a block and the corresponding $\mathbb{W}$, the whole block is loaded into memory (loading stage), and the walk updates (executing stage) are totally in memory without disk I/Os. 
Under on-demand load mode, only a portion of the block is loaded in the format of CSR segmentation (loading stage). The walk updates incur new disk I/Os when newly activated vertices are extended (executing stage). 
Compared with the full-load mode, the loading stage of on-demand load might be shorter, and the executing stage may get longer because of new disk I/Os. Therefore, we treat the two phases together as a whole to estimate the cost. 

For a certain block $B$, let $N_v$ be the number of total vertices in the block, and $\mathbb{A}$ be the set of activated vertices. It is intuitive that when $|\mathbb{A}|$ is very close to $N_v$, it should be more efficient to process the block under full-load mode than on-demand-load mode. This is because under such circumstances loading an entire block is faster than $|\mathbb{A}|$ small I/Os, which accelerates the loading stage, and since there is no need to invoke I/Os when executing walk updates, the executing stage is also faster. In random walk tasks, it is very expensive to obtain the accurate $|\mathbb{A}|$ when $|\mathbb{W}|$ is large, so we use $|\mathbb{W}|$ to roughly estimate $|\mathbb{A}|$. 
Let $\eta = |\mathbb{W}| / N_v$, which roughly represents the ratio of vertices whose information is needed in the block $B$. Let $t_f$ and $t_o$ be the total time of the loading and executing stage under full-load mode and on-demand load mode, respectively. Empirical studies on datasets in Table~\ref{tab:datasets} show that there exists an $\eta_0$ such that in general:
\begin{equation}
    \left\{
    \begin{array}{rcl}
    t_f > t_o & \text{if} & \eta > \eta_0;\\
    t_f < t_o & \text{if} & \eta < \eta_0.
    \end{array}\right.
\end{equation}
We further find out that $t_f$-$\eta$ follows a linear regression model $t_f = \alpha_f \eta + b_f$ for each block, and $t_o$-$\eta$ follows $t_o = \alpha_o \eta$ when $\eta < \eta_0$. Here $b_f$ means the cost of loading stage in full-load mode, and no intercept exists in $t_o$-$\eta$ model because no separated loading is needed when $\mathbb{W} = \emptyset$ under on-demand-load mode.

\subsubsection{Model Training and Learning of Thresholds}
To train the parameters $\alpha_f, b_f, \alpha_o$, we run the task twice to get the running log. Full-load mode is used for ancillary blocks in the first run, while in the second run, the on-demand load mode is used. We get the $t_f$-$\eta$ and $t_o$-$\eta$ running logs, and use these data to train $\alpha_f$, $b_f$ and $\alpha_o$. 
Therefore, we calculate $\eta_0 = \frac{b_f}{\alpha_o - \alpha_f}$, and use $\eta_0$ as the loading mode switching threshold for the ancillary block. Specifically, if $\eta > \eta_0$, full-load mode is used, and on-demand-load mode is used otherwise. 

\section{Implementation} \label{sec:implementation}
\begin{figure}
    \centering
    \includegraphics[width=8cm]{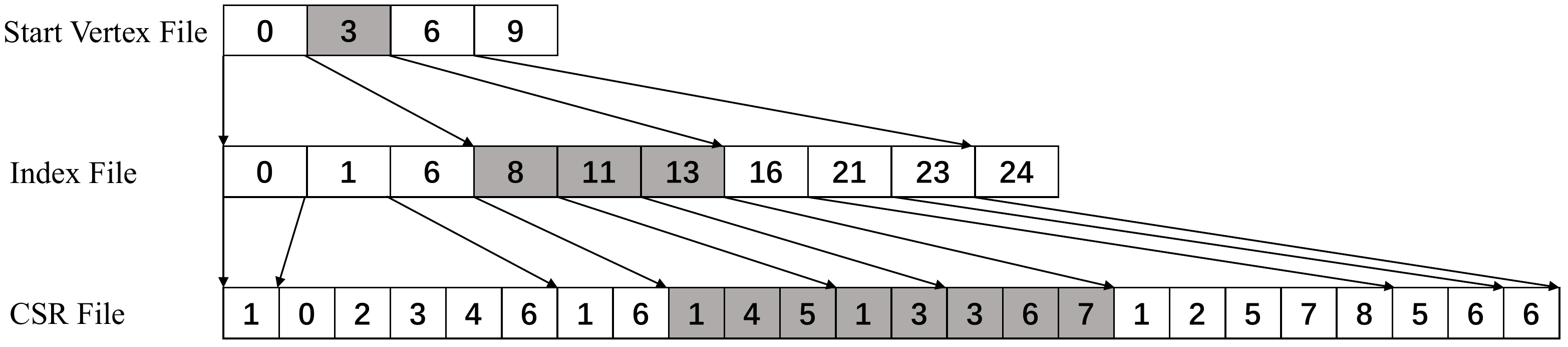}
    \caption{The CSR storage of the graph in Figure \ref{fig:example}. The grey cells represent the information of block 1.}
    \label{fig:CSR}
\end{figure}

In this section, we present the implementation details of \sys.
The graph is stored in the \textit{Compressed Sparse Row (CSR)} format and is sequentially partitioned into blocks according to the IDs of the vertices. Figure \ref{fig:CSR} is an example of the graph data organization in disk. The \textit{Start Vertex File} indicates the start vertex of each block, and the index file specifies the start offset of the neighbors of a vertex. The Start Vertex File is read into memory at the very beginning when \sys~ starts, so to record the IDs of the start vertex of each block.
 
\subsection{Walk Representation}
\begin{figure}
    \centering
    \includegraphics[width=8cm]{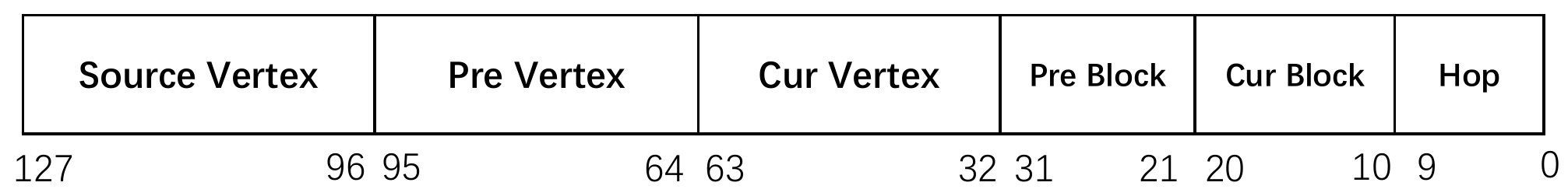}
    \caption{Walk representation.}
    \label{fig:walk-representation}
\end{figure}

To efficiently track the second-order random walks, we encode each walk with 128 bits shown in Figure \ref{fig:walk-representation}. \textit{Source Vertex} and \textit{Pre Vertex} are the start vertex and previous vertex of the walk, and \textit{Cur Vertex} is the offset of the current vertex in its residing blocks. \textit{Pre Block} and \textit{Cur Block} indicate the previous and current residing block of the walk, and \textit{Hop} marks how many steps the walk has already moved. With the help of this walk encoding, \sys~ can process graphs with at most 4.3 trillion vertices, partitioned to no more than 1024 blocks. For a single walk, the maximum moving steps supported is 1024.

\subsection{Graph Partition}
The default graph partition method used in \sys~ is a sequential partition, which partitions the vertices in their ID order to different blocks and ensures that all blocks fit the pre-defined block size. \sys~ also supports custom graph partition, and users have to provide the corresponding edge list file and the block file. The edge list file is composed of all the edges in the graph, and the block file indicates the block each vertex belongs to.

In this paper, we also study the impact of different graph partition methods on the overall performance. For random walk tasks, updating as many steps as possible in a block is preferred. Increasing the density of a block provides help since there are more edges between vertices in this block, and the walks are more likely to stay inside the block comparing with the sparse block. 
METIS~\cite{metis} is a popular and classical graph partition algorithm that aims to increase the density of each block. 
The influence of METIS partition to \sys~is discussed in Section \ref{sec:exp-metis}. We find out that customized graph partition methods like METIS can help improve the efficiency of \sys, but they take expensive time to pre-partition the whole graphs, and may face the out-of-core problem. While the sequential partition method is very fast. 
\subsection{Parallelism} \label{sec:parallelism}
During the processing of each bucket, multi-threads are executed in parallel. This brings the challenge of dealing with critical sections. As shown in Line \ref{line:move2bucket} in Algorithm \ref{alg:walk-processing}, the bucket-extending operation is used to append certain walks to buckets to be executed later in the current time slot. This forms a critical section because multi-threads may access the same bucket at the same time. Simple mutex that only allows one thread to write the bucket at one time brings in excessive synchronization overhead. To avoid this, we allocate buffers for each thread to store walks that meet the bucket-extending condition. Before executing a bucket, the buffer of each thread corresponding to this bucket is scanned, and walks in them are finally appended to the bucket. The performance of running the Node2vec random walk model on Twitter saves 40\% of overall time by applying such optimization compared to the mutex solution. Such an idea is also used when associating walks with blocks, where we set multi-walk buffers for each walk pool to store walks produced by multi-threads.
\section{Experiments}
In this section, we report the experimental results on \sys. We first show the end-to-end performance of \sys, by comparing with two variants implemented on GraphWalker. Then we focus on the two main techniques proposed in this paper, respectively. For the bi-block execution engine, we compare its performance with a plain bucket model, which is implemented based on a simple bucket-based walk collection idea. For the learning-based block loading model, we verify the improvement of I/O utilization and execution efficiency by using this model in \sys.
{We then test the impact of different graph partition methods.  
We also show the parameter sensitivity of \sys, including the variation of walk distribution and block size. Moreover, we evaluate the efficiency of \sys~on a set of synthetic datasets with different graph distributions. Finally, we discuss the applicability of \sys~for first-order random walks.}

\subsection{Experimental Settings}
\label{sec:expr_settings}
All experiments are run on a server with 2 Intel Xeon(R) Gold 5220 CPU @ 2.20GHz and 377 GB memory. The graph data is stored on an SSD. Without specific clarification, each experiment is run in parallel, and the number of threads is set to 72. 

\textbf{Datasets.}~{We use 6 datasets in our experiments: LiveJournal (LJ) \footnote{https://snap.stanford.edu/data/soc-LiveJournal1.html}, Twitter (TW)\footnote{https://old.datahub.io/dataset/twitter-social-graph-www2010}, Friendster (FR)\footnote{https://snap.stanford.edu/data/com-Friendster.html}, UK200705 (UK)\footnote{http://law.di.unimi.it/webdata/uk-2007-05/}, Kron29, which is a synthetic graph generated by Graph500 kronecker\footnote{https://graph500.org/} and CrawlWeb (CW)\footnote{http://webdatacommons.org/hyperlinkgraph/index.html}.} The statistics and partitions are listed in Table \ref{tab:datasets}. 
In the table, CSR Size indicates the storage cost to store graphs in CSR format, and Text Size is the size of the dataset stored in text format as an edge list. 
{Block Size is a parameter set manually by the user.} 
Block Number is the number of blocks that a graph is partitioned to, giving the block size. All graphs are processed into undirected.

\begin{table}[t]
\footnotesize
\begin{tabular}{cccccccc}
\hline
\textbf{Graph} & \textbf{|V|} & \textbf{|E|}         & \textbf{\begin{tabular}[c]{@{}c@{}}Text\\ Size\end{tabular}} & \textbf{\begin{tabular}[c]{@{}c@{}}CSR\\ Size\end{tabular}} & \textbf{\begin{tabular}[c]{@{}c@{}}Block\\ Size\end{tabular}} & \textbf{\begin{tabular}[c]{@{}c@{}}Block\\ Number\end{tabular}} & \textbf{\begin{tabular}[c]{@{}c@{}}Edge-\\ Cut\end{tabular}} \\ \hline
LiveJournal    & 4.8M         & 85.7M                & 1.2GB                                                        & 364MB                                                       & 20000KB                                                       & 17                                                              & 76.51\%                                                      \\ \hline
Twitter        & 41.7M        & 2.4B                 & 37GB                                                         & 9.3GB                                                       & 512MB                                                         & 18                                                              & 89.36\%                                                      \\ \hline
Friendster     & 65.6M        & 3.6B                 & 58GB                                                         & 14GB                                                        & 512MB                                                         & 27                                                              & 91.43\%                                                      \\ \hline
UK200705       & 105M         & 6.6B                 & 6.6B                                                         & 26GB                                                        & 1GB                                                           & 25                                                              & 32.49\%                                                      \\ \hline
Kron29         & 277M         & 33.7B                & 497GB                                                        & 128GB                                                       & 10GB                                                          & 13                                                              & 92.66\%                                                      \\ \hline
{CrawlWeb}       & {3.6B}         & {226B} & {4.6TB}                                                        & {864GB}                                                       & {100GB}                                                         & {9}                                                               &       -                                                       \\ \hline
\end{tabular}
\caption{{Graph datasets and partition information.}}
\label{tab:datasets}
\end{table}

\textbf{Second-order random walk models.} We use Node2vec models in our experiments. As Equation \ref{equ:node2vec} shows, the Node2vec model is controlled by two hyper-parameters $p$ and $q$. In this paper, we mainly focus on I/O performance {improvement}, so we set $p$ and $q$ to 1 as a representation for comparison. 


\textbf{Benchmarks.} Different benchmarks define different random walk tasks on the second-order random walk models. In our experiments, we choose two different benchmarks:
\begin{enumerate}
    \item Random walk generation using the Node2vec model {(RWNV)}. Following the random walk sampling approach in Node2vec, every node in a graph samples a set of random walks with a fixed length. Here we use the same parameter settings from the original work~\cite{Grover2016}, and sample 10 walks per-vertex with walk length of 80.
    \item PageRank Query using the Node2vec Model {(PRNV)}. Given a query node $v$, we run the second-order random walk with restart to estimate the second-order PageRank value~\cite{Wu2016}. The decay factor is 0.85, the maximum length is 20, and the total sample size is $4|V|$. In addition, we randomly choose 10 to 100 query nodes for each dataset according to their data size. 
\end{enumerate}

\textbf{Baselines}: Since there is no existing systems designed for out-of-core second-order random walk processing, we take the following two implementations as baselines:
\begin{enumerate}
    \item Second-Order GraphWalker (SOGW). The naive solution implemented on GraphWalker, which retrieves the previous vertex information directly from the disk as small vertex I/Os. We set the number of blocks in memory to 2 to make the memory cost equal to that of the other two methods. That is, if the block going to be load is already in memory, then no block loading is required. The block replacement strategy is the same as that of GraphWalker.
    \item Second-order GraphWalker with Static Cache (SGSC). A static vertex cache whose size is the same as the block size is set in memory. Before the execution starts, we tally the out-degrees of all vertices in the graph and store the top-$k$ vertices such that the degree sum is no less than the maximum edge number in one block according to the block size. There is no item replacement of the vertex cache during the execution.
\end{enumerate}

\subsection{End-to-End Performance}\label{sec:entire-performance}

\begin{figure}[t]
\subfigure[{Efficiency of RWNV.}]{
\includegraphics[width=0.4\textwidth]{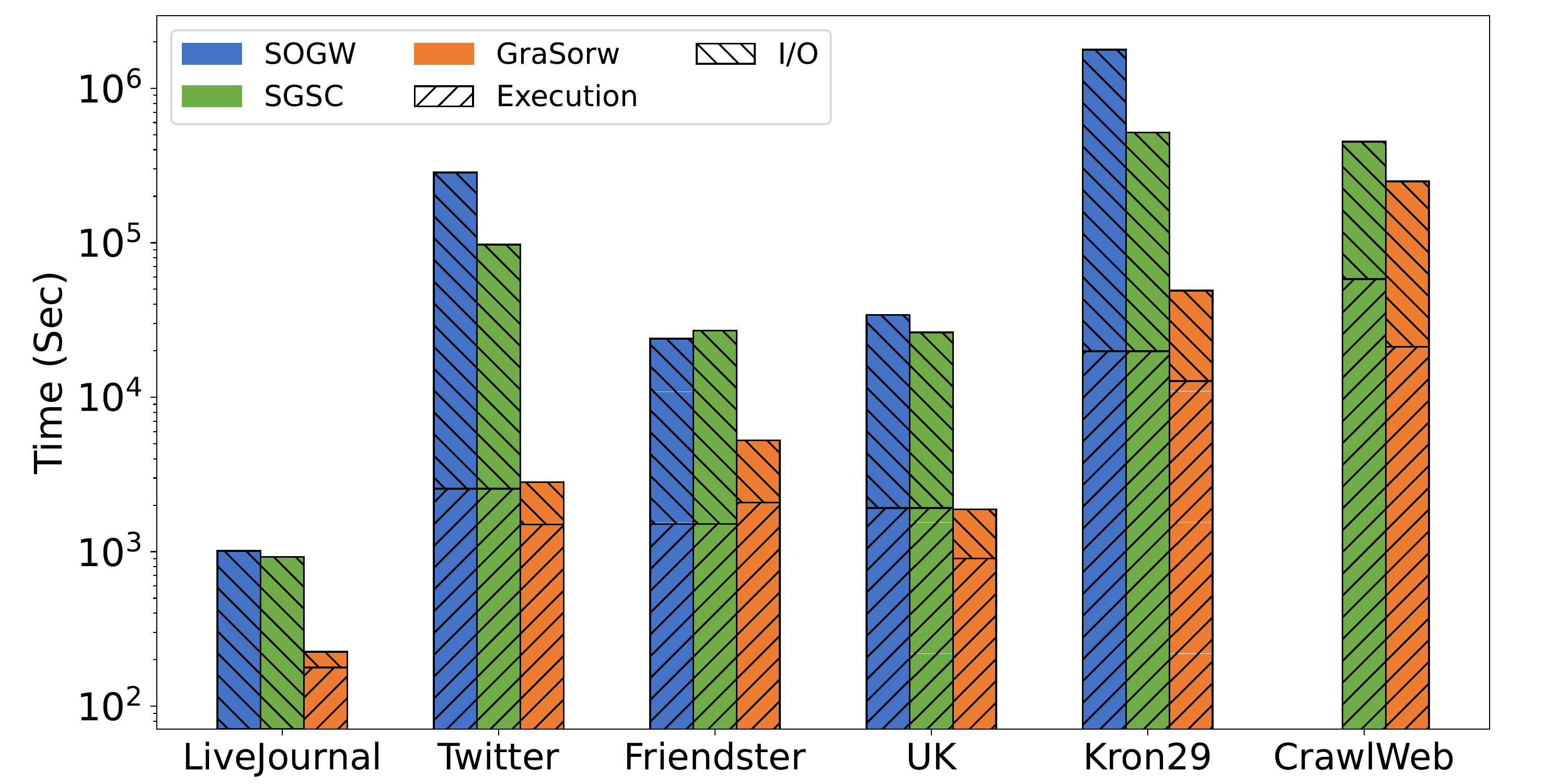}
\label{fig:entire-performance-nvnv}}
\subfigure[{Average query time of PRNV}]{\includegraphics[width=0.4\textwidth]{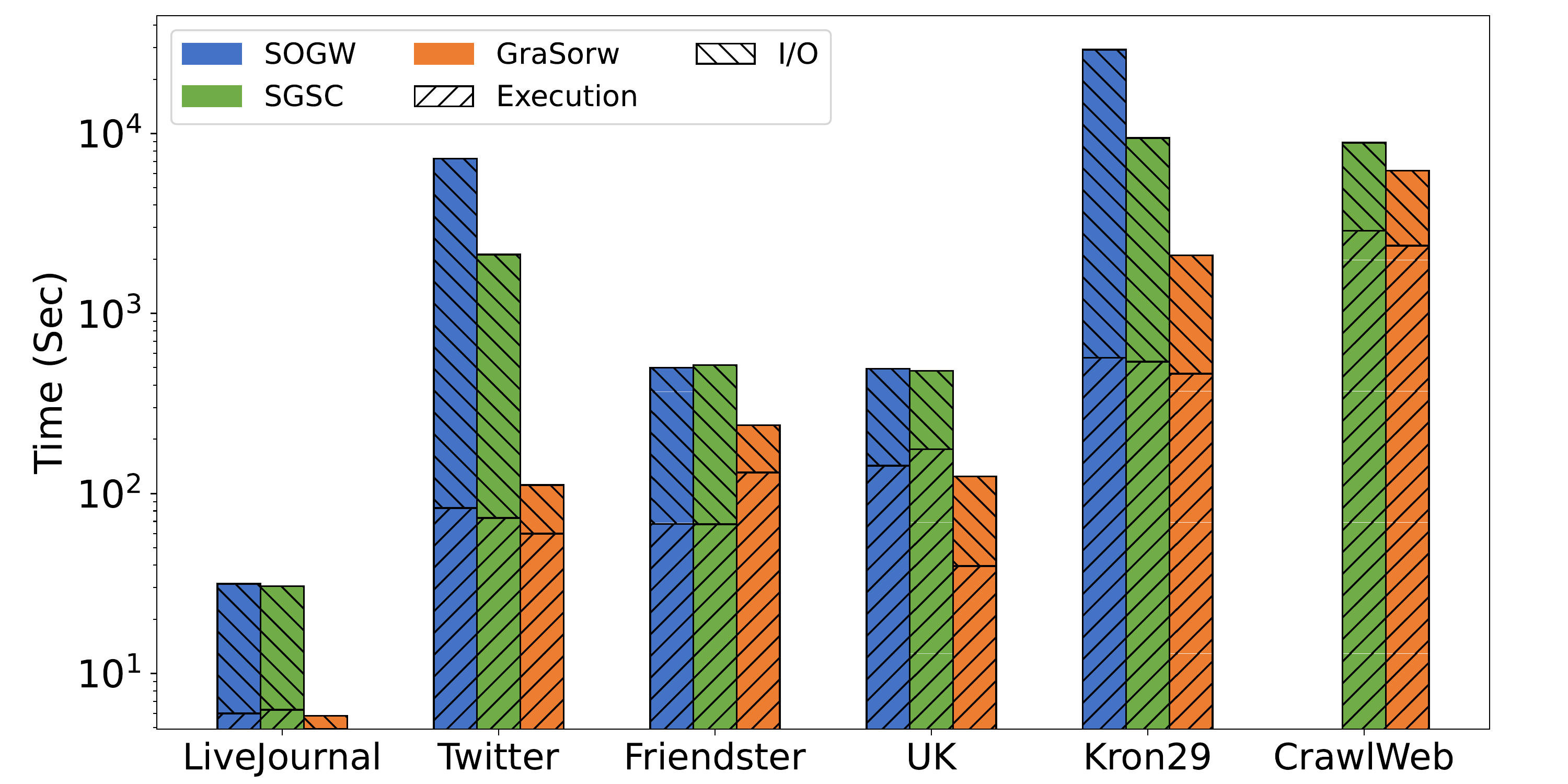}\label{fig:entire-performance-pprnv}}
\caption{End-to-end performance comparison. The y-axis shows the time cost in log-scale. On LiveJournal, the execution time of SOGW and SGSC is around 70 seconds, and it is too small to be visualized in the log-scale setting.}
\label{fig:e2e_performance}
\end{figure}
We first evaluate the overall performance of \sys~compared to the two baseline systems, SOGW and SGSC. 
Due to the inefficiency of SOGW and SGSC, they cannot finish the tasks with standard parameters in reasonable time constraints when processing large graphs except LiveJournal. 
In this paper, we estimate their costs on graphs except for LiveJournal as below:
According to the empirical studies in GraphWalker~\cite{Wang2020}, the total time increases linearly with the walk length when it is not possible to put the whole graph in memory. Besides, we find that when running second-order random walk tasks on SOGW, since the previous vertex information should be retrieved from disk, which accounts for most of the time as shown in Figure \ref{fig:io-percentage}, the total time also increases linearly with the number of walks. Therefore, we shorten the walk length for RWNV and start fewer walks from a vertex for both RWNV and PRNV. After obtaining the cost of the small-scale task, we estimate the cost by multiplying the corresponding coefficients. In addition, all results of \sys~are obtained by running the complete task.


Figure \ref{fig:e2e_performance} presents the results of RWNV and PRNV on various graphs. 
{The results of SOGW on CrawlWeb are missing, because the small-scale task used to estimate the total time cannot finish in two days. With this lower bound of the small-scale task execution time, we estimate that SOGW cannot finish the complete task in two weeks for both RWNV and PRNV. }
For SGSC, the time of the vertex cache initialization is included in I/O time.
Among three systems, we can see that \sys~achieves the best performance in both tasks on all these graphs. 
In particular, on Twitter, SOGW takes more than two days to finish the RWNV task, while \sys~ only takes 47 minutes, which achieves 95$\times$ speed up. On a larger graph that occupies hundreds of Gigabytes in CSR format such as Kron29, performing second-order random walk tasks is much more challenging, as the traditional disk-based methods cost about 20 days, evaluated by SOGW. Fortunately, with the help of \sys, such a task can be finished in half a day, which is much more reasonable. {On CrawlWeb, which takes almost 900GB of memory in CSR format, \sys~still achieves the best efficiency for both tasks, with a speedup of 1.81$\times$ for RWNV and 1.43$\times$ for PRNV, compared to \cache.} In most graphs, the SGSC is slightly faster than SOGW, as a result of the existence of static vertex cache in memory, which makes it possible to retrieve the information of some important vertices from memory rather than by invoking vertex I/Os.
{However, SGSC takes more time to run such a task on Friendster. One possible reason is that for Friendster, the cache hit rate in \cache~ is low so that the time of initiating the static vertex cache is longer than the time saved from its benefits. 
From the result comparison between \cache~ and \sys, we can see that for fixed memory size, rather than leverage the memory space to store as many large-degree vertices as possible, it is more efficient to use the memory to load blocks (i.e., ancillary blocks) with the triangular bi-block scheduling.} Overall, \sys~ has achieved {1.81$\times$ to 95$\times$} performance improvement in RWNV task, and {1.43$\times$ to 19.1$\times$} improvement in PRNV task. 

In the figure, we also present the time cost breakdown for each result, visualized as \textit{Execution} time and \textit{I/O} time. We see that the I/O time cost for each task on all graphs has decreased significantly in \sys. \sys~ reduces the I/O overhead most on Twitter, increasing efficiency by 213$\times$ in RWNV and 138$\times$ in PRNV, {comparing with SOGW.} In SOGW, the expensive I/O cost comes from the massive light vertex I/Os, while in \sys, with the help of buckets and the ancillary block, these vertex I/Os used to retrieve the information of the previous vertex of walks are converted into block I/Os, which are more efficient. 

\begin{table*}[t]
\small
\begin{tabular}{cc|cccc|cccc}
\hline
\multirow{2}{*}{\textbf{graph}}  & \multirow{2}{*}{\textbf{Method}} & \multicolumn{4}{c|}{\textbf{RWNV}}                                                                                                                                                                                                                                                      & \multicolumn{4}{c}{\textbf{PRNV}}                                                                                                                                                                                                                                                       \\ \cline{3-10} 
                                 &                                  & \textbf{\begin{tabular}[c]{@{}c@{}}Wall\\ Time (s)\end{tabular}}   & \textbf{\begin{tabular}[c]{@{}c@{}}Execution\\ Time (s)\end{tabular}} & \textbf{\begin{tabular}[c]{@{}c@{}}Block I/O\\ Number\end{tabular}} & \textbf{\begin{tabular}[c]{@{}c@{}}Block I/O\\ Time(s)\end{tabular}} & \textbf{\begin{tabular}[c]{@{}c@{}}Wall\\ Time (s)\end{tabular}}   & \textbf{\begin{tabular}[c]{@{}c@{}}Execution\\ Time (s)\end{tabular}} & \textbf{\begin{tabular}[c]{@{}c@{}}Block I/O\\ Number\end{tabular}} & \textbf{\begin{tabular}[c]{@{}c@{}}Block I/O\\ Time(s)\end{tabular}} \\ \hline
\multirow{2}{*}{\textbf{LJ}}     & \textbf{PB}                      & 332                                                                & 189                                                                   & 13584                                                               & 90                                                                   & 9.8                                                                & 5.7                                                                   & 38                                                                  & 3                                                                    \\ \cline{2-10} 
                                 & \textbf{Bi-Block}                & \textbf{\begin{tabular}[c]{@{}c@{}}175\\ ($53\%$)\end{tabular}}    & \textbf{\begin{tabular}[c]{@{}c@{}}100\\ ($53\%$)\end{tabular}}       & \textbf{\begin{tabular}[c]{@{}c@{}}6299\\ ($46\%$)\end{tabular}}    & \textbf{\begin{tabular}[c]{@{}c@{}}42\\ ($46\%$)\end{tabular}}       & \textbf{\begin{tabular}[c]{@{}c@{}}5.8\\ ($6\%$)\end{tabular}}     & \textbf{\begin{tabular}[c]{@{}c@{}}3.6\\ ($63\%$)\end{tabular}}       & \textbf{\begin{tabular}[c]{@{}c@{}}21\\ ($56\%$)\end{tabular}}      & \textbf{\begin{tabular}[c]{@{}c@{}}2\\ ($64\%$)\end{tabular}}        \\ \hline
\multirow{2}{*}{\textbf{TW}}     & \textbf{PB}                      & 6868                                                               & 1905                                                                  & 15779                                                               & 4463                                                                 & 249.8                                                              & 83.5                                                                  & 419                                                                 & 138.2                                                                \\ \cline{2-10} 
                                 & \textbf{Bi-Block}                & \textbf{\begin{tabular}[c]{@{}c@{}}2960\\ ($43\%$)\end{tabular}}   & \textbf{\begin{tabular}[c]{@{}c@{}}1198\\ ($63\%$)\end{tabular}}      & \textbf{\begin{tabular}[c]{@{}c@{}}8090\\ ($51\%$)\end{tabular}}    & \textbf{\begin{tabular}[c]{@{}c@{}}1384\\ ($31\%$)\end{tabular}}     & \textbf{\begin{tabular}[c]{@{}c@{}}111.6\\ ($45\%$)\end{tabular}}  & \textbf{\begin{tabular}[c]{@{}c@{}}44.2\\ ($53\%$)\end{tabular}}      & \textbf{\begin{tabular}[c]{@{}c@{}}255\\ ($61\%$)\end{tabular}}     & \textbf{\begin{tabular}[c]{@{}c@{}}43.9\\ ($32\%$)\end{tabular}}     \\ \hline
\multirow{2}{*}{\textbf{FR}}     & \textbf{PB}                      & 14526                                                              & 3982                                                                  & 34117                                                               & 9743                                                                 & 439.9                                                              & 103.6                                                                 & 964                                                                 & 283.5                                                                \\ \cline{2-10} 
                                 & \textbf{Bi-Block}                & \textbf{\begin{tabular}[c]{@{}c@{}}6716\\ ($46\%$)\end{tabular}}   & \textbf{\begin{tabular}[c]{@{}c@{}}3217\\ ($81\%$)\end{tabular}}      & \textbf{\begin{tabular}[c]{@{}c@{}}18550\\ ($54\%$)\end{tabular}}   & \textbf{\begin{tabular}[c]{@{}c@{}}2882\\ ($3\%$)\end{tabular}}      & \textbf{\begin{tabular}[c]{@{}c@{}}240\\ ($55\%$)\end{tabular}}    & \textbf{\begin{tabular}[c]{@{}c@{}}102.7\\ ($99\%$)\end{tabular}}     & \textbf{\begin{tabular}[c]{@{}c@{}}581\\ ($6\%$)\end{tabular}}      & \textbf{\begin{tabular}[c]{@{}c@{}}94.3\\ ($33\%$)\end{tabular}}     \\ \hline
\multirow{2}{*}{\textbf{UK}}     & \textbf{PB}                      & 20707                                                              & 4143                                                                  & 29309                                                               & 16043                                                                & 554.1                                                              & 102.1                                                                 & 659                                                                 & 379.6                                                                \\ \cline{2-10} 
                                 & \textbf{Bi-Block}                & \textbf{\begin{tabular}[c]{@{}c@{}}3789\\ ($18\%$)\end{tabular}}   & \textbf{\begin{tabular}[c]{@{}c@{}}744\\ ($18\%$)\end{tabular}}       & \textbf{\begin{tabular}[c]{@{}c@{}}10039\\ ($34\%$)\end{tabular}}   & \textbf{\begin{tabular}[c]{@{}c@{}}2596\\ ($16\%$)\end{tabular}}     & \textbf{\begin{tabular}[c]{@{}c@{}}146.5\\ ($26\%$)\end{tabular}}  & \textbf{\begin{tabular}[c]{@{}c@{}}32\\ ($30\%$)\end{tabular}}        & \textbf{\begin{tabular}[c]{@{}c@{}}312\\ ($47\%$)\end{tabular}}     & \textbf{\begin{tabular}[c]{@{}c@{}}81.0\\ ($21\%$)\end{tabular}}     \\ \hline
\multirow{2}{*}{\textbf{Kron29}} & \textbf{PB}                      & 133491                                                             & 24312                                                                 & 19592                                                               & 104962                                                               & 5793.3                                                             & 827.0                                                                 & 878                                                                 & 4728                                                                 \\ \cline{2-10} 
                                 & \textbf{Bi-Block}                & \textbf{\begin{tabular}[c]{@{}c@{}}49694\\ ($37\%$)\end{tabular}}  & \textbf{\begin{tabular}[c]{@{}c@{}}12738\\ ($52\%$)\end{tabular}}     & \textbf{\begin{tabular}[c]{@{}c@{}}11608\\ ($59\%$)\end{tabular}}   & \textbf{\begin{tabular}[c]{@{}c@{}}34024\\ ($32\%$)\end{tabular}}    & \textbf{\begin{tabular}[c]{@{}c@{}}2102.5\\ ($36\%$)\end{tabular}} & \textbf{\begin{tabular}[c]{@{}c@{}}366.9\\ ($44\%$)\end{tabular}}     & \textbf{\begin{tabular}[c]{@{}c@{}}520\\ ($59\%$)\end{tabular}}     & \textbf{\begin{tabular}[c]{@{}c@{}}1582.3\\ ($34\%$)\end{tabular}}   \\ \hline
\multirow{2}{*}{\textbf{{CrawlWeb}}} & {\textbf{PB}}                      & {911114}                                                             & {316320}                                                                & {6384}                                                                & {568576}                                                               & {39649}                                                              & {22296}                                                                 & {100}                                                                 & {12309.4}                                                              \\ \cline{2-10} 
                                 & \textbf{{Bi-Block}}                & \textbf{\begin{tabular}[c]{@{}c@{}}{249529}\\($27\%$)\end{tabular}} & \textbf{\begin{tabular}[c]{@{}c@{}}{21206}\\($7\%$)\end{tabular}}      & \textbf{\begin{tabular}[c]{@{}c@{}}{2624}\\($41\%$)\end{tabular}}    & \textbf{\begin{tabular}[c]{@{}c@{}}{228256}\\($40\%$)\end{tabular}}   & \textbf{\begin{tabular}[c]{@{}c@{}}{6218.1}\\($16\%$)\end{tabular}} & \textbf{\begin{tabular}[c]{@{}c@{}}{892.8}\\($4\%$)\end{tabular}}      & \textbf{\begin{tabular}[c]{@{}c@{}}{45}\\($45\%$)\end{tabular}}      & \textbf{\begin{tabular}[c]{@{}c@{}}{3772.6}\\($31\%$)\end{tabular}}   \\ \hline
\end{tabular}
\caption{I/O efficiency of different execution engines. Wall time is the total running time of the task, and it is decomposed into execution time, block I/O time and other overheads such as walk initiating and walk loading. Execution time is the cost of walk updating. The {percentages} in parentheses are the ratio of the cost of Bi-Block to the one of PB, respectively.} 
\label{tab:io-efficiency}
\end{table*}
\subsection{The I/O-Efficiency of Bi-Block Execution Engine}
\label{sec:expr-engine}
Here we compare the I/O efficiency of two execution engines in \sys, the plain bucket engine (PB) and the bi-block execution engine (Bi-Block). The former organizes walks in buckets without the triangular bi-block scheduling strategy and the skewed walk storage. 
In the plain bucket engine, walks are associated with their current blocks, and the current walks are split into buckets according to their previous blocks. There are also two block slots in memory called the current block and the ancillary block, yet no triangular bi-block scheduling is used, where the schedule of the ancillary block starts from $b_0$ to $b_{N_b - 1}$. We use the state-aware block scheduling strategy proposed by GraphWalker to schedule the current block. 

The results of the two engines are shown in Table \ref{tab:io-efficiency}. 
The wall time with the bi-block engine is $18\%$ to $53\%$ of the one with the plain bucket engine for RWNV, and {$16\%$} to $60\%$ for PRNV. The performance improvement is more significant on larger graphs such as {UK, Kron29 and CrawlWeb}, which is more than 60\%. Next, we deeply compare the block I/Os and execution time of the two engines.

\textbf{Block-I/O comparison.} We first focus on the block I/O overhead of two engines. The block I/O number in the bi-block execution engine is only 34\% to 59\% of the one in the plain bucket model for RWNV, and {45\%} to 61\% for PRNV, respectively. This is consistent with the theoretical analysis that the triangular bi-block scheduling strategy approximately cuts half of the block I/Os, according to Equation \ref{eq:bi-block}. Concretely, the reason for block I/O reduction is twofold. First, during each time slot, half of the ancillary block loading whose ID is less than the current block is saved. Second, the current block loading stops on $b_{N_b - 2}$, so one block loading for the current block is saved in each time slot. 
Another observation is that when processing large graphs except for LJ, the block I/O time of the bi-block engine is reduced to {16\% to {40\%}} on both tasks, which is less than the reduction factor (i.e., 50\%) of block I/O number. 
{One reason is that some expensive random I/Os of loading blocks in the plain bucket engine are converted to sequential block I/Os in the bi-block execution engine during the block scheduling.}
In the plain bucket engine, after loading the current block, the loading of ancillary blocks starts from $b_0$, which incurs a random block I/O, while in \sys, the loading starts from the next block to the current block, which is sequential. 

\textbf{Execution time comparison.} The execution time of the bi-block execution engine also decreases compared to the one of the plain bucket execution engine. For example, on UK, the execution time of the bi-block execution engine decreases to 18\% for RWNV and 30\% for PRNV, {and on CrawlWeb, such decrease reaches 7\% and 4\% for RWNV and PRNV, respectively.} One reason is that the bi-block execution engine reduces the thread management overhead by reducing the number of block I/Os in a time slot. Concretely, in \sys~the current walks are executed in parallel, and the system needs to manage the threads, like the initiating, destroying. 
In both two engines, each loading of the ancillary block corresponds to a bucket execution. In the bi-block execution engine, the number of bucket execution is only half of that in the plain bucket engine, since the number of ancillary block I/Os has been reduced to around 50\% as discussed above. Therefore, the initiating and destroying overhead in thread management is decreased, leading to the decrease of total execution time. 
\begin{table*}[]
\footnotesize
\begin{tabular}{cc|cccc|cccccc}
\hline
\multirow{2}{*}{\textbf{Graph}} & \multirow{2}{*}{\textbf{Partition}} & \multicolumn{4}{c|}{\textbf{Pure Full Load}}                                                                                                                                                                                                                                     & \multicolumn{6}{c|}{\textbf{Learning-based Load}}                                                                                                                                                                                                                                                                                                                                                                                  \\ \cline{3-12} 
                                &                                     & \textbf{\begin{tabular}[c]{@{}c@{}}Wall\\ Time(s)\end{tabular}} & \textbf{\begin{tabular}[c]{@{}c@{}}Execution\\ Time(s)\end{tabular}} & \textbf{\begin{tabular}[c]{@{}c@{}}Block I/O\\ Time(s)\end{tabular}} & \textbf{\begin{tabular}[c]{@{}c@{}}Block\\ I/O Num\end{tabular}} & \textbf{\begin{tabular}[c]{@{}c@{}}Wall\\ Time(s)\end{tabular}} & \textbf{\begin{tabular}[c]{@{}c@{}}Execution\\ Time(s)\end{tabular}} & \textbf{\begin{tabular}[c]{@{}c@{}}Block I/O\\ Time(s)\end{tabular}} & \textbf{\begin{tabular}[c]{@{}c@{}}Block\\ I/O Num\end{tabular}} & \textbf{\begin{tabular}[c]{@{}c@{}}On-demand\\ I/O Time(s)\end{tabular}} & \textbf{\begin{tabular}[c]{@{}c@{}}On-demand\\ I/O Num\end{tabular}} \\ \hline
\textbf{TW}                     & \textbf{Seq}                        & 3434                                                            & 1317                                                                 & 1689                                                                 & 9936                                                             & 3229                                                            & 1266                                                                 & 1419                                                                 & 8224                                                             & 61                                                                       & 1714                                                                 \\ \hline
\textbf{TW}                     & \textbf{METIS}                      & 2829                                                            & 1039                                                                 & 1541                                                                 & 7540                                                             & 2465                                                            & 1053                                                                 & 1056                                                                 & 5145                                                             & 96                                                                       & 2168                                                                 \\ \hline
\textbf{UK}                     & \textbf{Seq}                        & 4798                                                            & 662                                                                  & 3705                                                                 & 13587                                                            & 2992                                                            & 1467                                                                 & 749                                                                  & 2650                                                             & 332                                                                      & 10628                                                                \\ \hline
\textbf{UK}                     & \textbf{METIS}                      & 1856                                                            & 98                                                                   & 1044                                                                 & 3751                                                             & 1165                                                            & 166                                                                  & 294                                                                  & 998                                                              & 38                                                                       & 2558                                                                 \\ \hline
\end{tabular}
\caption{The performance of different loading methods with different partitions for the RWNV task.}
\label{tab:block-loading-rwnv}
\end{table*}

\subsection{The Effectiveness of Learning-based Block Loading}\label{sec:exp-block-loading}

In this experiment, we first describe the efficiency of \sys~ when using the learning-based block loading model, then analyze the improvement of I/O utilization. 

\begin{figure}[t]
    \centering
    \subfigure[RWNV]{
        \label{subfig:block-loading-rwnv}
        \includegraphics[width=0.22\textwidth]{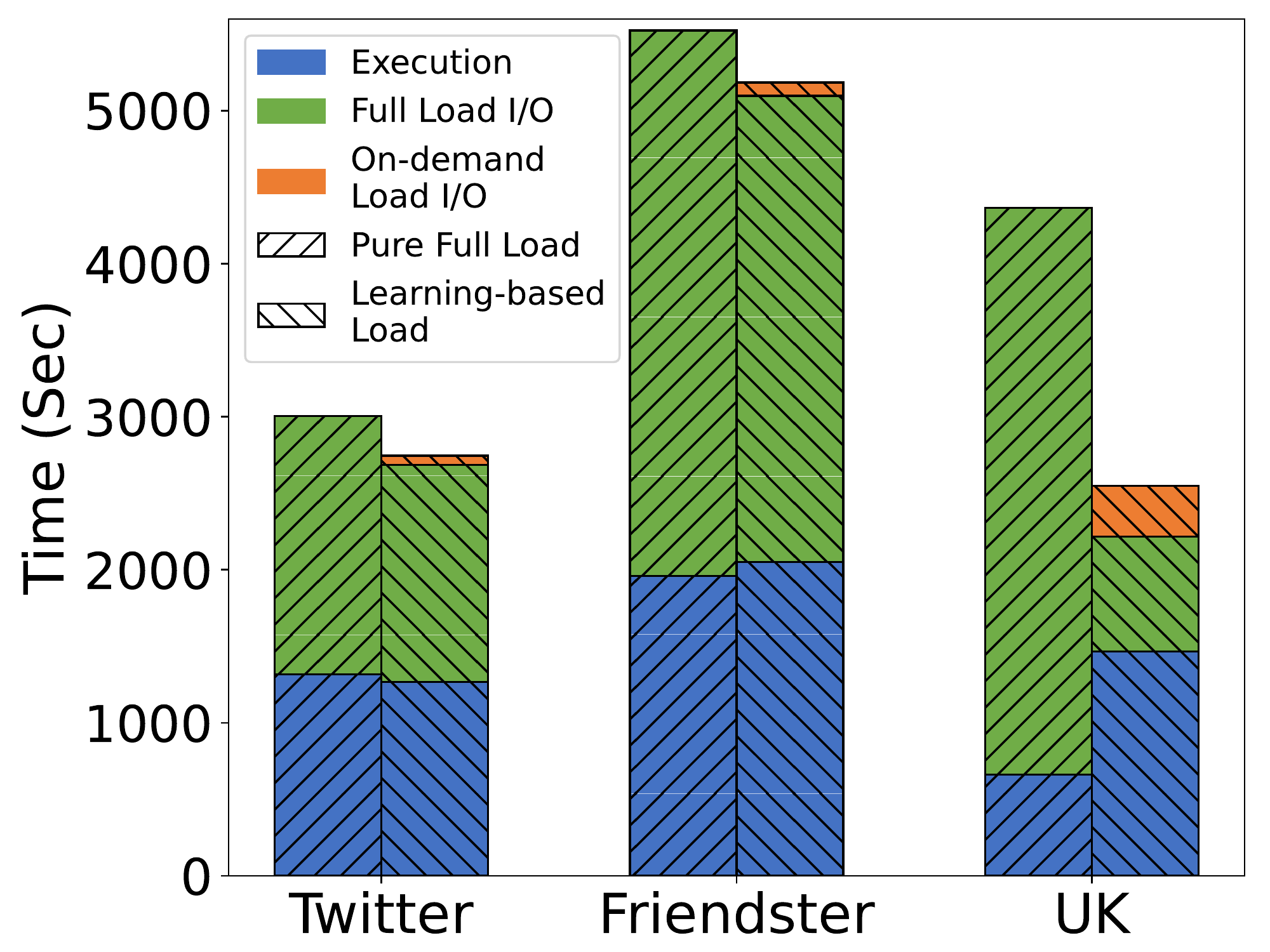}
    }
    \subfigure[PRNV]{
        \label{subfig:block-loading-prnv}
        \includegraphics[width=0.22\textwidth]{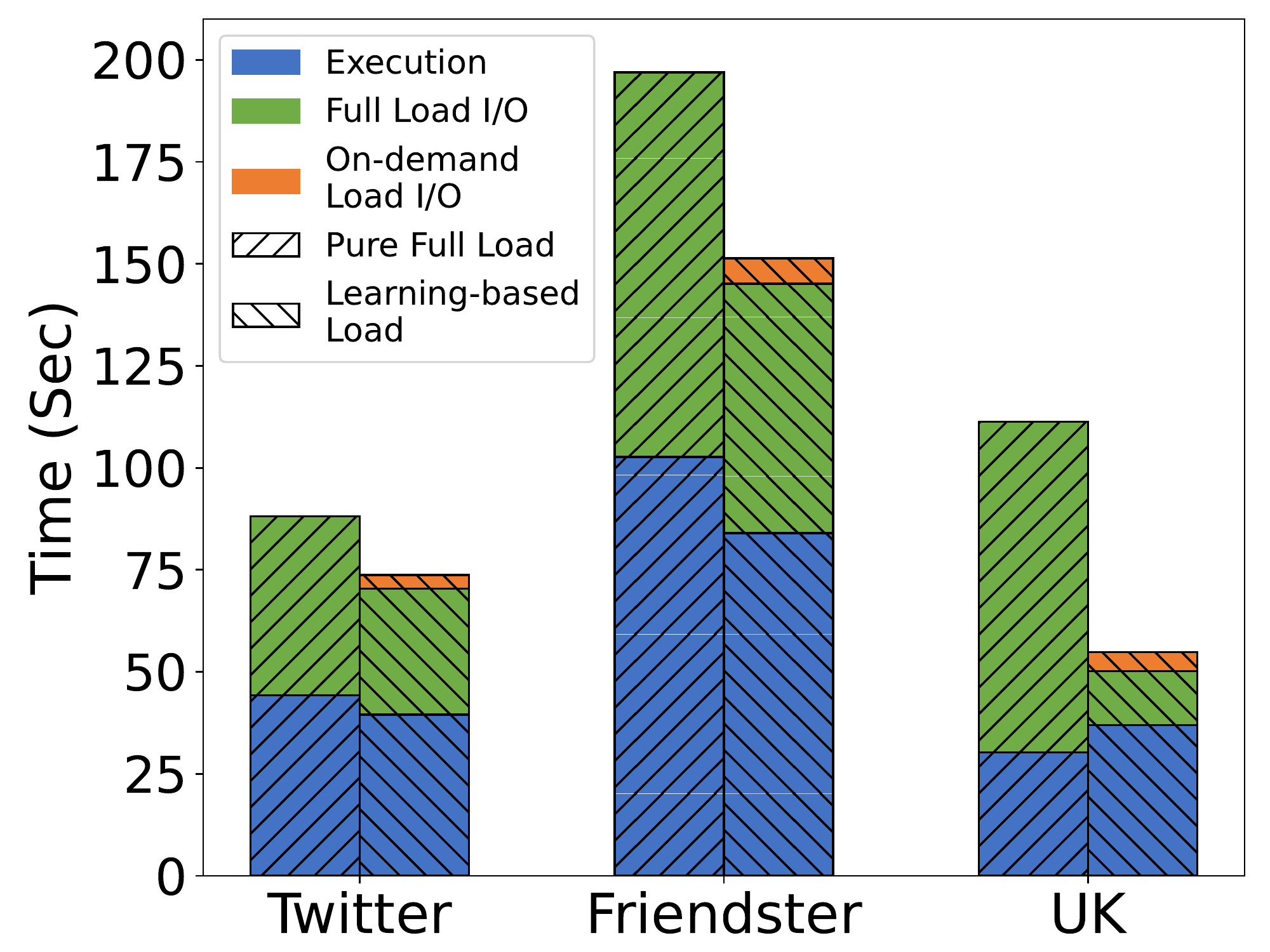}
    }
    \caption{Performance of the learning-based block loading model on various datasets.}
    \label{fig:block-loading}
\end{figure}

\textbf{Efficiency}. The overall performance of using the learning-based block loading model is shown in Figure \ref{fig:block-loading}. 
The execution time is the time cost of updating walks. Under pure full load mode, the execution time does not include any I/O costs, while using the learning-based block loading method, the execution time includes some I/O costs which is incurred by the on-demand block loading method because the on-demand block loading method might bring in random vertex I/O during walk processing to get the vertex information that has not been loaded at the beginning. 
Therefore, in most of the results, the execution time of the learning-based method is longer than that of the full-load mode. For example, in Figure \ref{subfig:block-loading-rwnv}, the execution time increases 805 seconds in graph UK compared to the pure full-load mode. However, the total time by using the learning-based block loading model is less because of the reduction of block I/Os (see results of sequential partition on UK in Table~\ref{tab:block-loading-rwnv}).
To be concrete, the difference between block I/O time in pure full load mode and the sum of block I/O time and on-demand load I/O time in learning-based block loading model is 2624 seconds (see results of sequential partition on UK in Table~\ref{tab:block-loading-rwnv}), which is much greater than the increase of execution time. 
Such trade-off is leveraged by the learning-based model described in Section \ref{sec:io-demand}. 

\begin{figure}
    \centering
    \subfigure[Full block loading.]{
        \label{subfig:FL-AB}
        \includegraphics[width=0.22\textwidth]{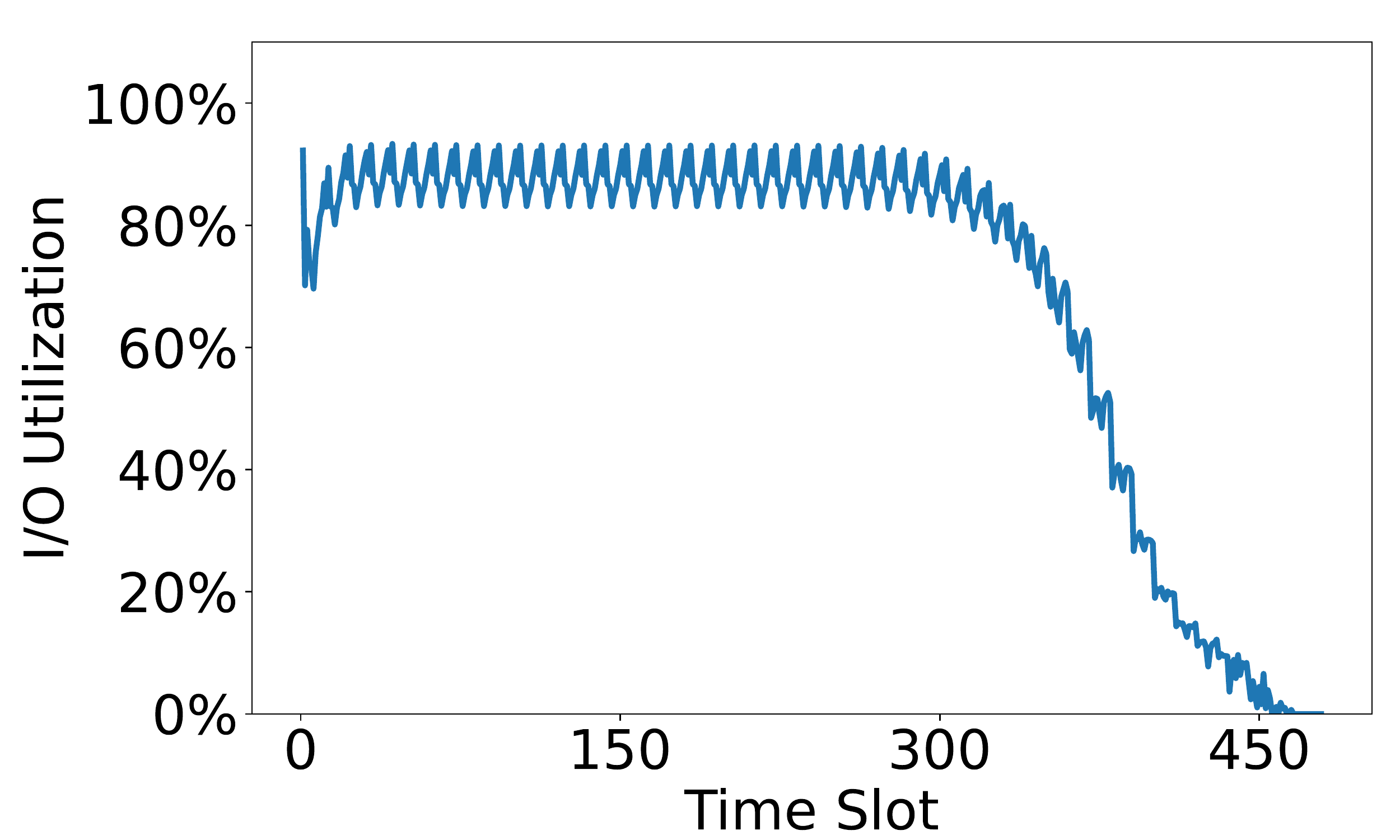}
    }
    \subfigure[Learning-based block loading.]{
        \label{subfig:OL-AB}
        \includegraphics[width=0.22\textwidth]{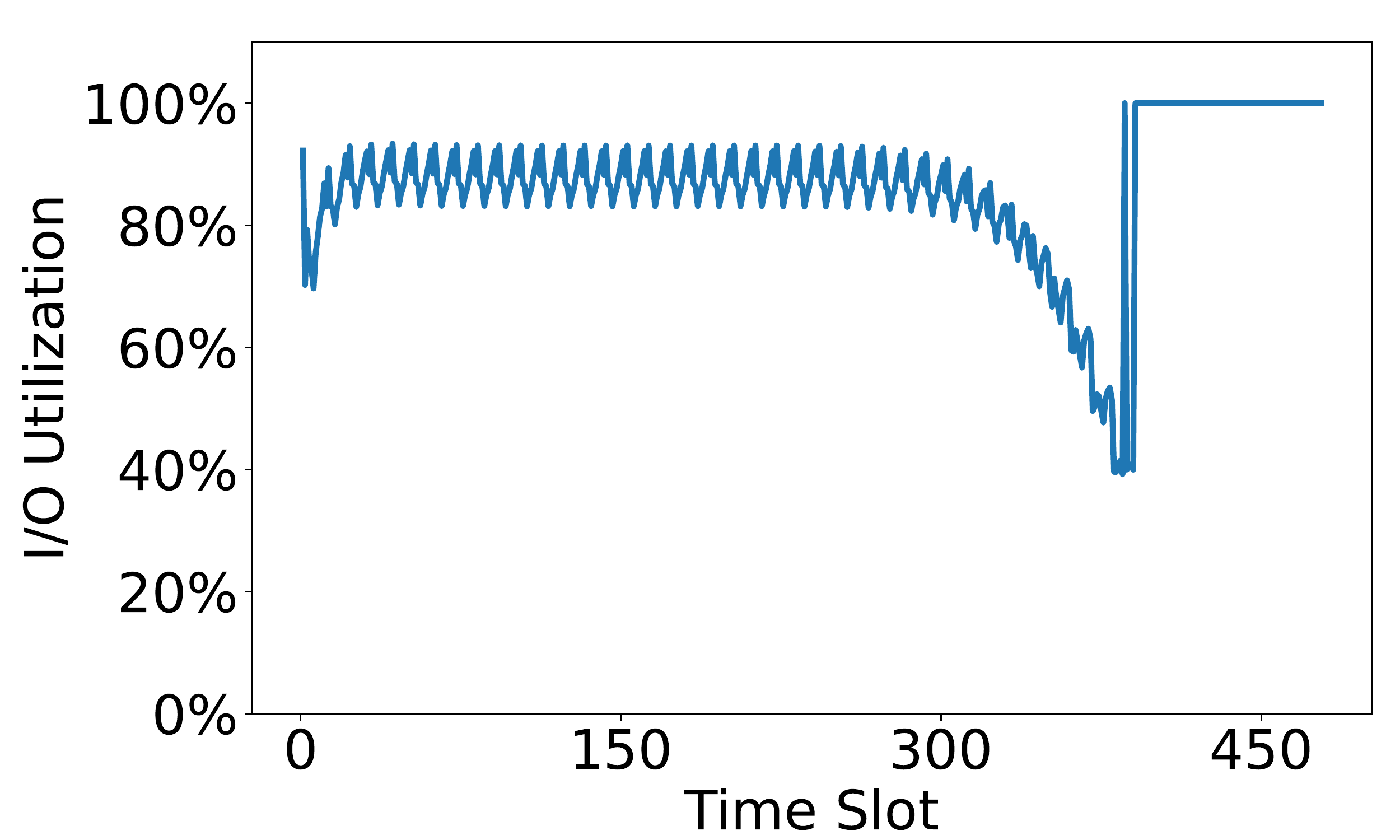}
    }
    \caption{I/O-utilization with different block loading methods on Twitter.}
    \label{fig:io-utiliazation}
\end{figure}

\textbf{I/O-Utilization}. We then discuss the I/O utilization with the two block loading methods. We take the I/O utilization of a specific block (e.g., block 10) in Twitter when it is loaded as an ancillary block as an example, and the results of other blocks in Twitter or other graphs are similar. Figure \ref{fig:io-utiliazation} shows the results under the pure full-load and learning-based block loading model. The I/O utilization is tallied after the execution of the corresponding bucket, {and the x-axis represents the time slot of each block loading}. The block I/O-utilization remains stable\footnote{There are periodic fluctuations during the plateau. This is caused by the different current blocks corresponding to the loaded ancillary block. The peak occurs when the current block has more edges connected with the ancillary block. To be concrete, during the plateau, each pair of the current block and ancillary block forms a stable block I/O-utilization. The current block varies periodically from the $b_0$ to the prior block to the ancillary block, leading to periodic fluctuations of the I/O-utilization of the ancillary block.} around 0.87 in the first 300 loads, we call this part the \textit{plateau}. 
After the plateau, the block I/O-utilization under pure full-load mode decreases close to 0. In this period, many walks reach the termination condition, and less information is required to update the few remaining walks. Since the pure full-load method still loads the whole block into memory, it suffers from low I/O utilization. Our learning-based block loading model is aware of the decreasing of update walks and is able to switch to on-demand loading mode. The on-demand loading ensures 100\% of I/O utilization, and only a small portion of block is loaded into memory.

\begin{figure*}[t]
    \centering
    \subfigure[RWNV $(p = 4, q = 0.25)$]{
        \label{subfig:rwnv-4-0.25}
        \includegraphics[width=0.22\textwidth]{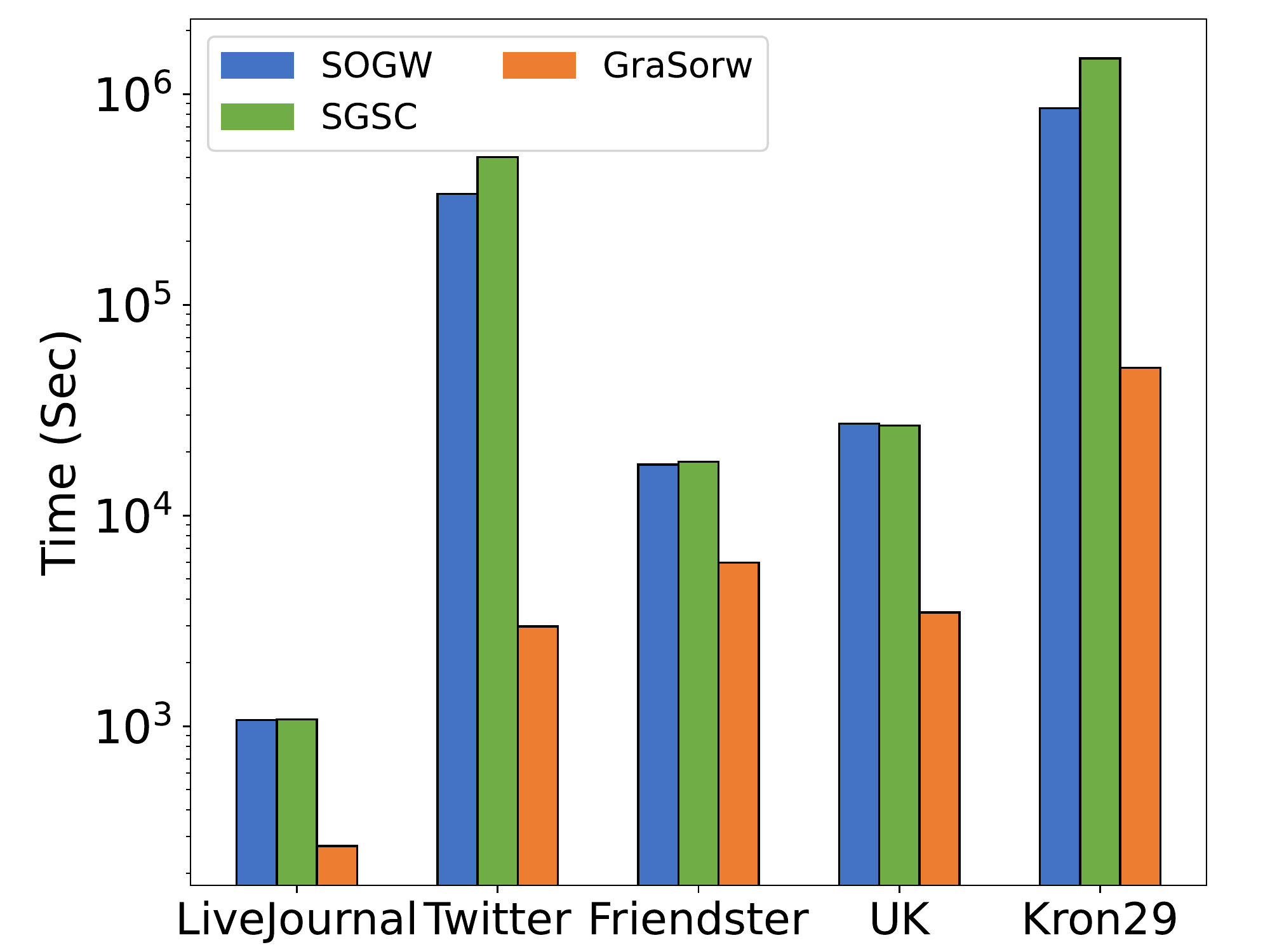}
    }
    \subfigure[RWNV $(p = 0.25, q = 4)$]{
        \label{subfig:rwnv-0.25-4}
        \includegraphics[width=0.22\textwidth]{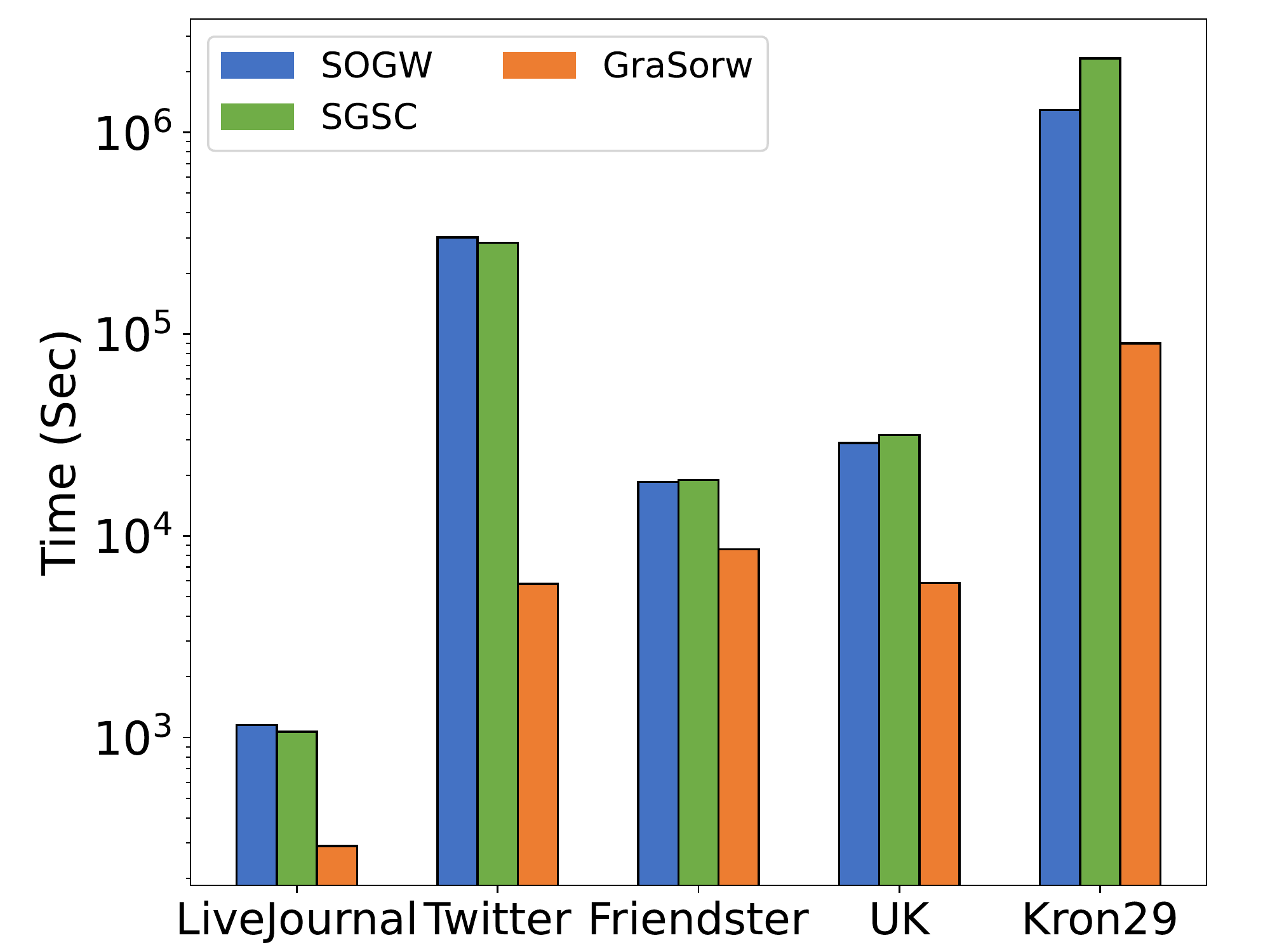}
    }
    \subfigure[PRNV $(p = 4, q = 0.25)$]{
        \label{subfig:prnv-4-0.25}
        \includegraphics[width=0.22\textwidth]{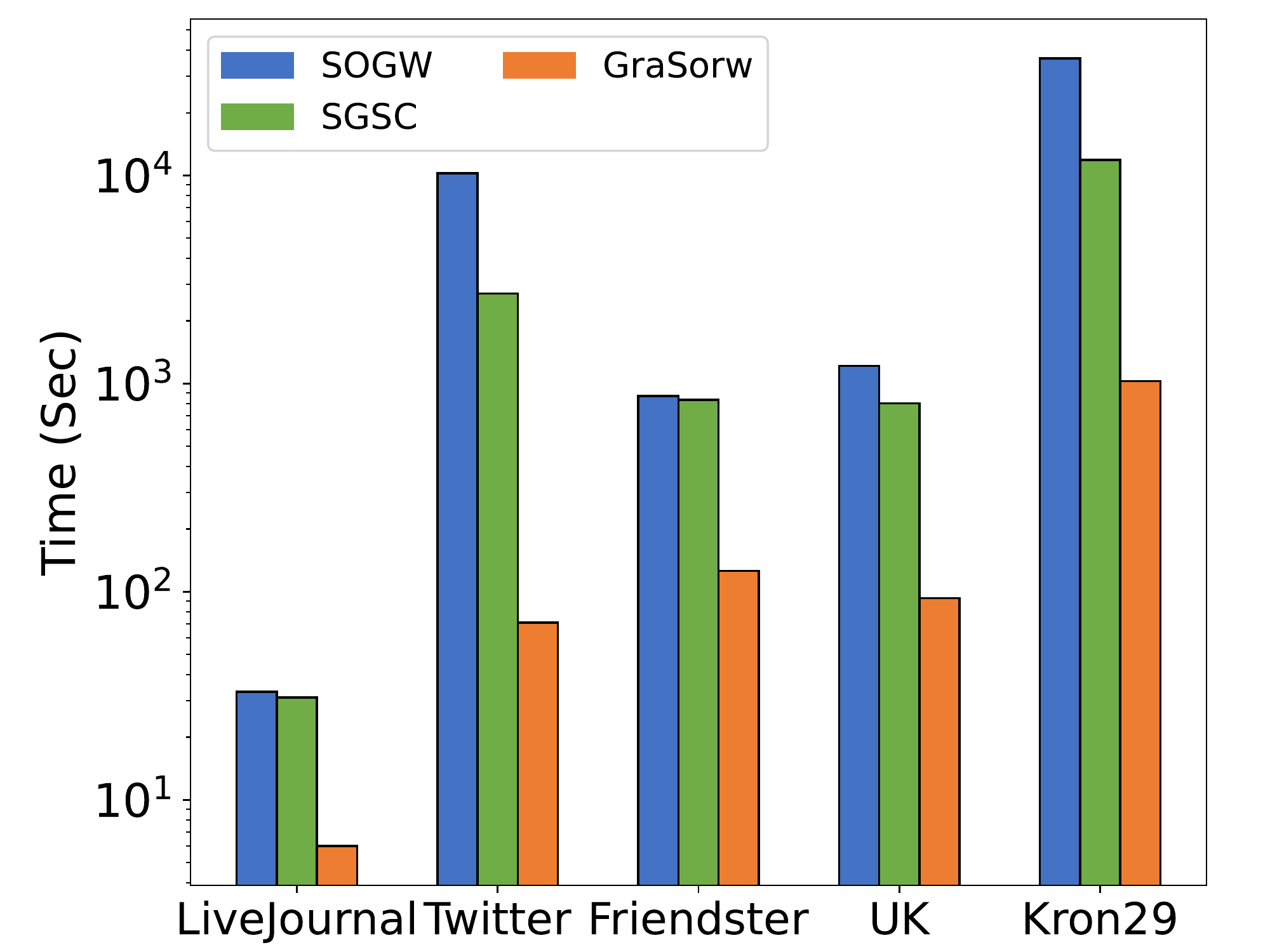}
    }
    \subfigure[PRNV $(p = 0.25, q = 4)$]{
        \label{subfig:prnv-0.25-4}
        \includegraphics[width=0.22\textwidth]{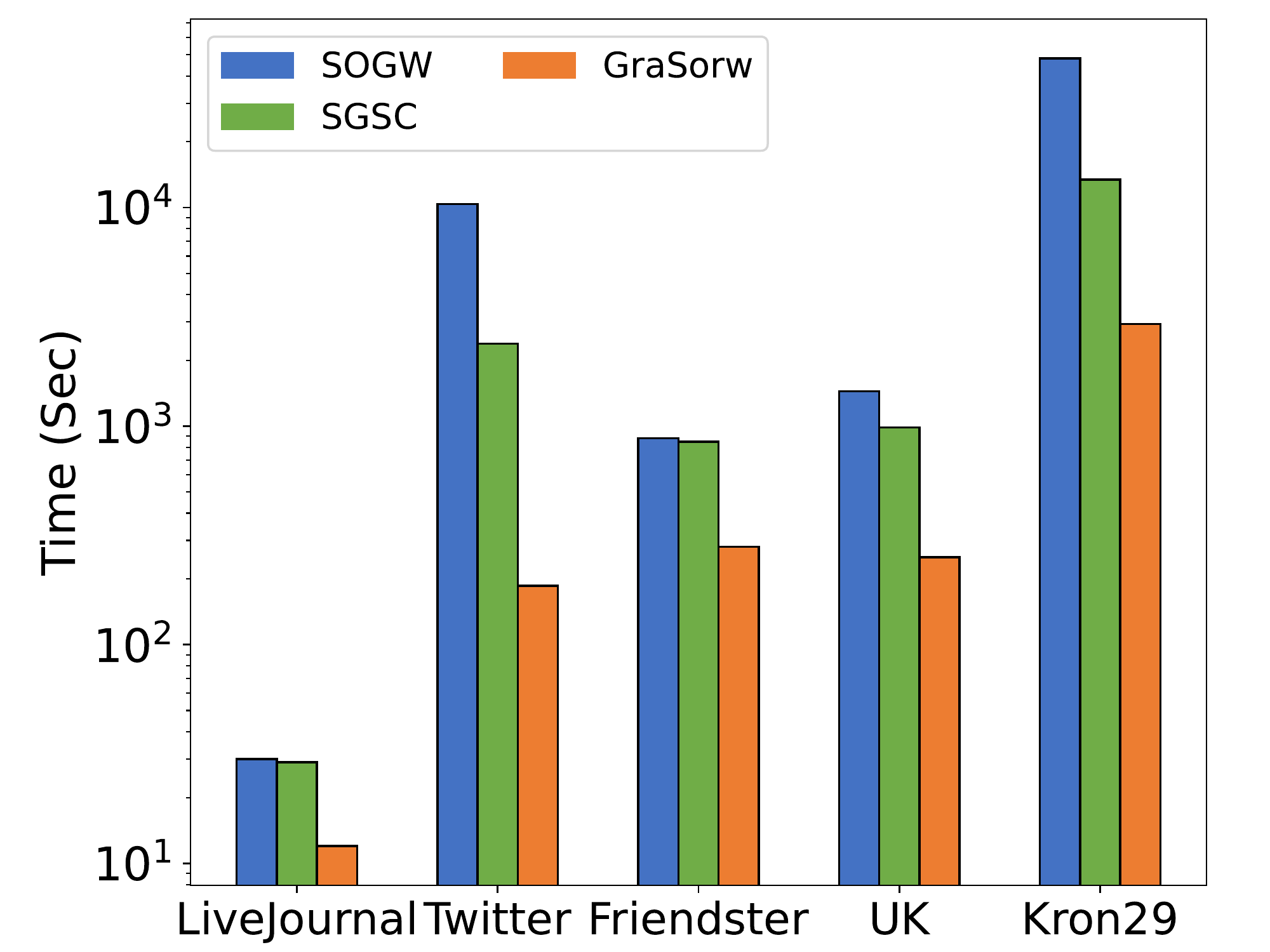}
    }
    \caption{{Wall time (log-scale) of RWNV and PRNV under different walk distributions.}}
    \label{fig:parameter-pq}
\end{figure*}
\begin{figure*}[t]
    \centering
    \subfigure[Twitter]{
        \label{subfig:TW}
        \includegraphics[width=0.3\textwidth]{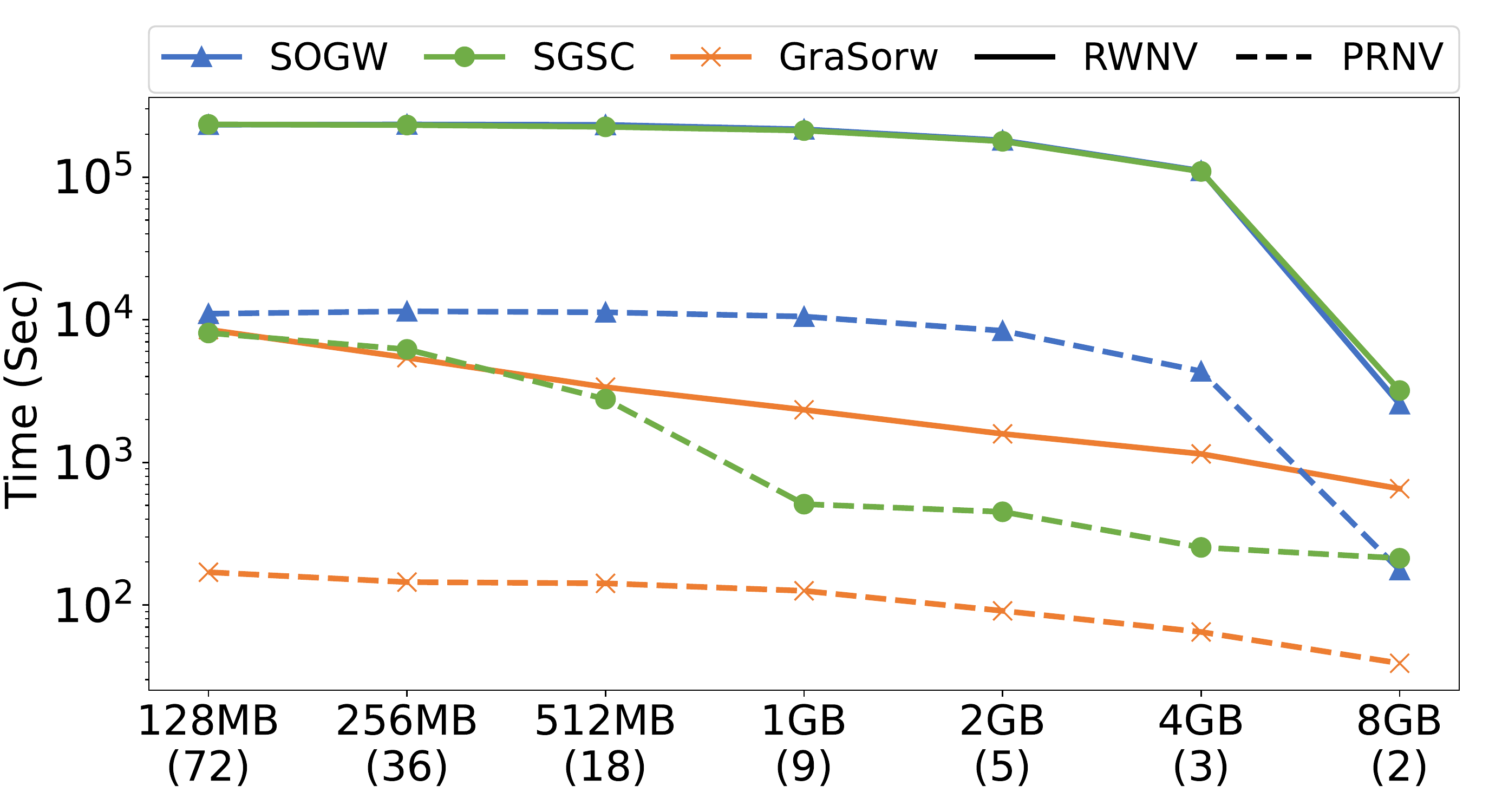}
    }
    \subfigure[Friendster]{
        \label{subfig:FR}
        \includegraphics[width=0.3\textwidth]{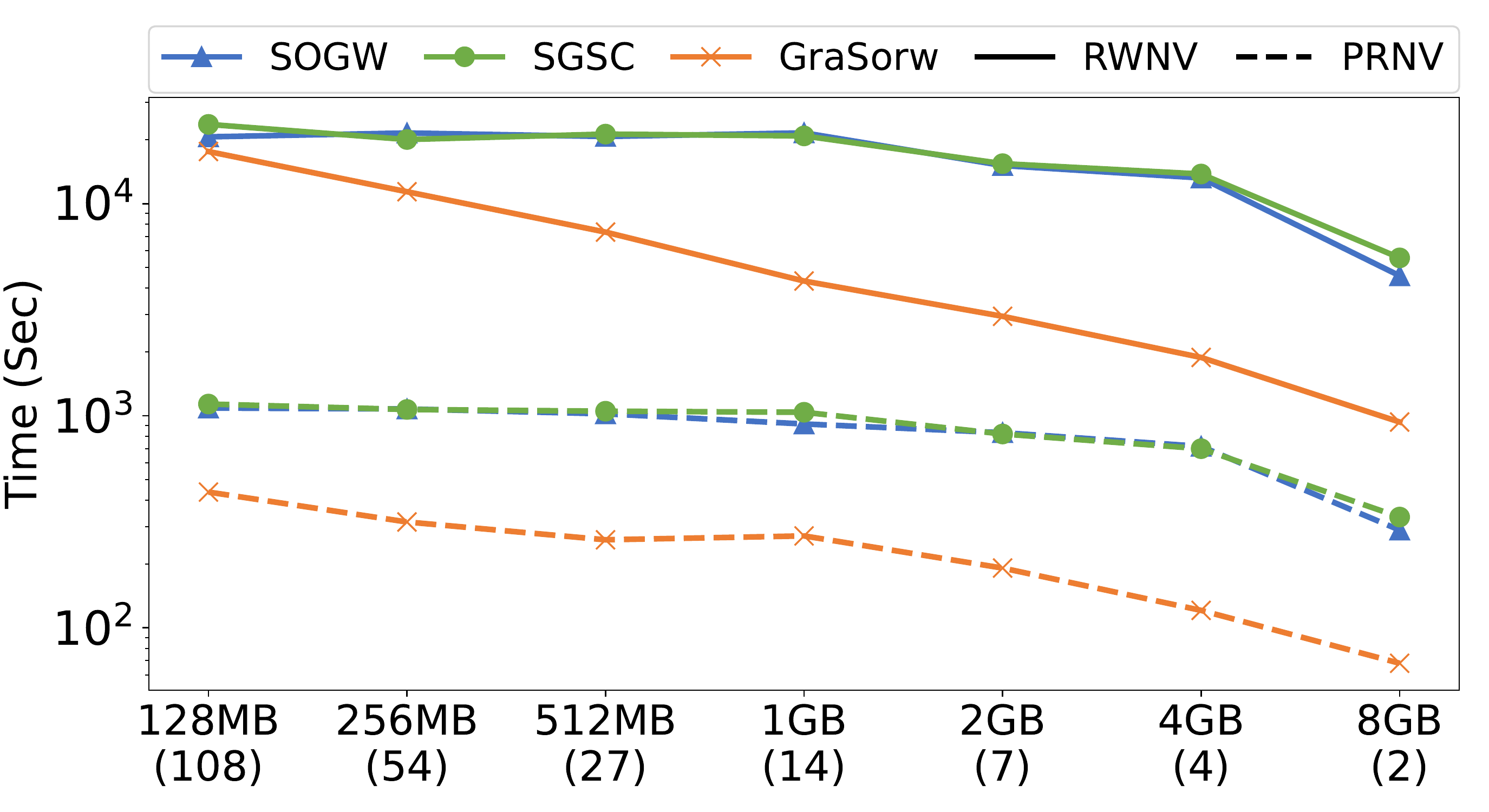}
    }
    \subfigure[{UK200705}]{
        \label{subfig:UK}
        \includegraphics[width=0.3\textwidth]{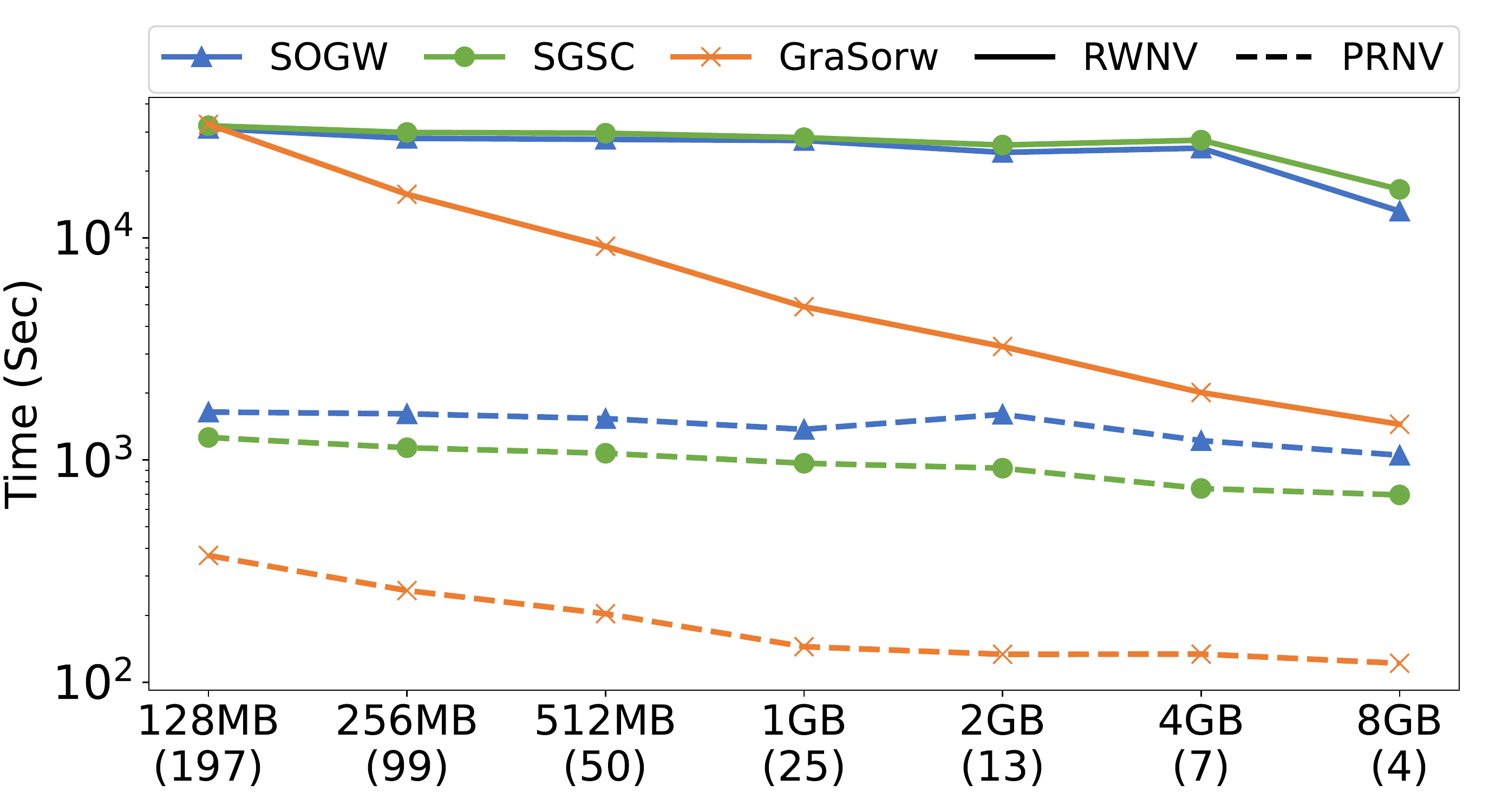}
    }
    \caption{{Wall time (log-scale) of RWNV and PRNV under different block sizes. In x-axis, the number in parentheses represents the number of partitioned blocks.}}
    \label{fig:parameter-blocksize}
\end{figure*}
\subsection{The Influence of Different Partitions} \label{sec:exp-metis}
As discussed in Section~\ref{sec:implementation}, different graph partition methods influence the overall performance. We compare the performance of \sys~under METIS partition and the default sequential partition with the RWNV task. 
Since METIS fails to partition graph FR and Kron29 in memory on our server, we only evaluate the performance of METIS partition on Twitter and UK, and the results of running RWNV task are shown in Table \ref{tab:block-loading-rwnv}. The partitioned block number is the same as that in sequential partitioning, and we ensure that the size of each block is roughly the same, where the biggest block is not larger than 1.03$\times$ the smallest one. We use the default k-way partition algorithm provided by METIS. For graph UK, the edge-cut of the METIS graph partition decreases significantly to $0.33\%$. For graph Twitter, the edge-cut is $55.14\%$. 

The block I/O number of METIS partition is reduced compared to that in sequential partition for both loading methods. Specifically, for UK, the number of block I/O dropped by 72\% under METIS partition in pure full load mode and 63\% in the learning-based load mode. This is because the density of blocks is increased, and walks are more likely to update inside the block than moving out of the block. Consequently, walks are able to move more steps forward during a time slot, and the block I/O number is reduced. Furthermore, according to the discussion in Section \ref{sec:expr-engine}, the decrease of block I/Os leads to the improvement of execution time as well.

{Another observation is that METIS partition improves the efficiency of \sys~when using the learning-based block loading model. Under sequential partition, the learning-based block loading model reduces 6\% of wall time in Twitter comparing with the pure full load method, while under METIS partition, the reduction reaches 13\%. 
This is caused by the decrease of edge-cut. For a graph partition with a lower edge-cut, most of the walks are able to reach the termination condition with fewer time slots because they tend to move forward inside the block. However, there are still few walks that tend to jump between different blocks, thus causing lots of block I/Os that have low I/O utilization under pure full-load mode. With the learning-based block loading model, these block I/Os can be completed by the on-demand loading method, thus increasing the I/O utilization and improving efficiency.}

\subsection{{Parameter Sensitivity}}
\subsubsection{Random Walk Distribution.}

{
To study the performance of \sys~under different walk distributions, we add sensitivity experiments with different $p$, $q$ of Node2vec, and the results of wall time are shown in Figure \ref{fig:parameter-pq}. 
We can see that \sys~achieves the least wall time (i.e., best efficiency) in all cases. For RWNV, since all of the vertices are activated, the static cache strategy in \cache~does not bring in significant improvement. In fact, on Kron29, the wall time of \cache~ is much longer, that is $1.7\times$ and $1.8\times$ that of SOGW under the two random walk distributions, respectively. On the other hand, \sys~is able to handle this situation and the wall time is only $6\%$ that of SOGW on Kron29 when $p=4,q=0.25$ and $7\%$ when $p=0.25,q=4$. 
For tasks such as PRNV in which only a few vertices are initially activated, \cache~saves more time comparing to that of RWNV. However, \sys~still achieves the best result. For instance, when executing PRNV ($p=4,q=0.25$) on Twitter, the wall time of \cache~is $26\%$ that of SOGW, while \sys~is only $0.7\%$. Moreover, on Friendster, \cache~only saves sightly $3\%$\textasciitilde$4\%$ of wall time of SOGW, the time saved by \sys~still achieves $86\%$ for PRNV ($p=4,q=0.25$) and $68\%$ for PRNV ($p=0.25,q=4$).}

\subsubsection{Block Size.}

{
We also study the performance of \sys~ with variation of block size and the number of blocks, and the results on three representative graphs are shown in Figure \ref{fig:parameter-blocksize}. 
Similar with previous tasks, there is no significant difference between SOGW and \cache~for RWNV, and for PRNV, the latter is only slightly faster. On the other hand,
\sys~consistently outperforms the baselines. The only exception happens when conducting RWNV on UK with block size set to 128MB. This is because when block size is small, the number of partitioned blocks of large graphs is big (197 in this case), and there are many ancillary block I/Os in \sys. 
Another observation is that as the block size increases, the advantage of \sys~becomes more and more obvious. For example, on Twitter, the maximum performance improvement by \sys~is achieved when the block size is set to 2GB, which is {114$\times$ and 91$\times$} speedups for RWNV and PRNV, respectively, comparing to the SOGW. 
The third observation is that when the block size reaches to 8GB, difference between \sys~and two baselines SOGW and \cache~ becomes less obvious. 
For example, on Twitter, the wall time of SOGW is about {97$\times$} slower than the one of \sys~when block size is 4GB (3 blocks in total), while it decreases to  4$\times$ slower when block size is enlarged to 8GB (2 blocks in total). 
This is because as the number of blocks decreases, walks can move more steps inside one block, and the number of walks crossing blocks becomes small, resulting fewer individual vertex I/Os. 
In summary, \sys~achieves significant time reduction under different block sizes and number of blocks.
}

\subsection{{The Performance of \sys~With Different Graph Distributions}}
{To study the performance of \sys~on graphs with different distributions, we generated 11 different graphs by considering three types of graph parameters, that are the skew parameters, density, and community structures. All these graphs are generated by using the popular Python package NetworkX\footnote{\url{https://networkx.org/documentation/stable/index.html}}. The information of each graph is shown in Table \ref{tab:synthetic-graph-info}. Note that these experiments focus on the influence of graph distributions to \sys, we did not generate very large graph, and the CSR sizes of generated graphs is controlled to be similar to the one of Twitter or LiveJournal. In the following, we discuss the experimental results of the three types one-by-one.}

\begin{table}[t]
\centering
\footnotesize
\scalebox{0.9}{
\begin{tabular}{ccccccccc}
\hline
                                              & \textbf{GraphName}  & \textbf{$|V|$} & \textbf{$|E|$} & \textbf{$\bar{D}$} & \textbf{\begin{tabular}[c]{@{}c@{}}Text\\ Size\end{tabular}} & \textbf{\begin{tabular}[c]{@{}c@{}}CSR\\ Size\end{tabular}} & \textbf{\begin{tabular}[c]{@{}c@{}}Block\\ Size\end{tabular}} & \textbf{\begin{tabular}[c]{@{}c@{}}\#Block\end{tabular}} \\ \hline
\multirow{3}{*}{\textbf{\tabincell{c}{Skew\\ness}}}            & \textbf{CirculantG} & 40M            & 1.6B           & 40                 & 26GB                                                         & 6.3GB                                                       & 512MB                                                         & 12                                                              \\ \cline{2-9} 
                                              & \textbf{RandomG}    & 40M            & 1.6B           & 40                 & 26GB                                                         & 6.3GB                                                       & 512MB                                                         & 12                                                              \\ \cline{2-9} 
                                              & \textbf{BASF}       & 40M            & 1.6B           & 40                 & 25GB                                                         & 6.3GB                                                       & 512MB                                                         & 12                                                              \\ \hline
\multirow{5}{*}{\textbf{Density}}             & \textbf{RandomG1}   & 100M           & 500M           & 5                  & 8.3GB                                                        & 2.7GB                                                       & 191MB                                                         & 10                                                              \\ \cline{2-9} 
                                              & \textbf{RandomG2}   & 10M            & 500M           & 50                 & 7.4GB                                                        & 1.9GB                                                       & 191MB                                                         & 11                                                              \\ \cline{2-9} 
                                              & \textbf{RandomG3}   & 1M             & 500M           & 500                & 6.5GB                                                        & 1.9GB                                                       & 191MB                                                         & 11                                                              \\ \cline{2-9} 
                                              & \textbf{RandomG4}   & 100K           & 500M           & 5000               & 5.5GB                                                        & 1.9GB                                                       & 191MB                                                         & 11                                                              \\ \cline{2-9} 
                                              & \textbf{RandomG5}   & 22360          & 500M           & 22359              & 5.2GB                                                        & 1.9GB                                                       & 191MB                                                         & 10                                                              \\ \hline
\multirow{3}{*}{\textbf{\tabincell{c}{Comm\\unity}}} & \textbf{SBM1}       & 42K            & 580M           & 13809              & 6.2GB                                                        & 2.2GB                                                       & 105MB                                                         & 21                                                              \\ \cline{2-9} 
                                              & \textbf{SBM2}       & 42K            & 1.06B          & 25238              & 12GB                                                         & 4.0GB                                                       & 195MB                                                         & 21                                                              \\ \cline{2-9} 
                                              & \textbf{SBM3}       & 42K            & 1.54B          & 36666              & 17GB                                                         & 5.8GB                                                       & 280MB                                                         & 21                                                              \\ \hline
\end{tabular}}
\caption{{Statistics of synthetic graphs.}
}
\label{tab:synthetic-graph-info}
\end{table}
{\textbf{Influence of skew parameters.} 
We generate a circulant graph\footnote{\url{https://networkx.org/documentation/stable/reference/generated/networkx.generators.classic.circulant_graph.html}} (CirculantG) and an Erdős-Rényi random network\footnote{\url{https://networkx.org/documentation/stable/reference/generated/networkx.generators.random_graphs.fast_gnp_random_graph.html\#networkx.generators.random_graphs.fast_gnp_random_graph}} (RandomG). We also use Barabási–Albert preferential attachment method to generate a scale-free network\footnote{\url{https://networkx.org/documentation/stable/reference/generated/networkx.generators.random_graphs.barabasi_albert_graph.html\#networkx.generators.random_graphs.barabasi_albert_graph}} (BASF), in which degrees of vertices follow power-law distribution. 
For the above three graphs, we make sure that they have the same number of vertices and edges. In other words, they have the same average degree, which is 40 in our experiments.
} 

{According to the results in Table~\ref{tab:syngraph-table}, \sys~achieves the best efficiency for all datasets. Concretely, \sys~achieves 4.6$\times$\textasciitilde6.1$\times$ speedup for RWNV, and 2.4$\times$\textasciitilde4.9$\times$ for PRNV, compared to SOGW. Because of the bucket-based walk management and the existence of the ancillary block, the massive individual vertex I/Os, which are random I/Os, are converted to few sequential block I/Os, and this makes \sys~be faster than baselines. The above results demonstrate that \sys~is able to achieve high-performance on graphs with different skew parameters.}

{\textbf{Influence of different densities.} To study the performance of \sys~under different graph densities, we generate five random graphs with average degree being 5 (RandomG1), 50 (RandomG2), 500 (RandomG3), 5000 (RandomG4) and 22359 (RandomG5), respectively. Note that fixing the number of total edges to 500 million, RandomG5 is the smallest graph that can be generated, i.e., the graph has the fewest vertices and is a complete graph.
These random graphs are generated with previously mentioned Erdős-Rényi random network model as well.}

{According to the results in Table~\ref{tab:syngraph-table}, \sys~achieves the best performance in most cases. Similar with experiments of skew parameters, the speedup is also achieved with the help of the conversion from massive light vertex I/Os to the sequential block I/Os. However, under different tasks (i.e, RWNV and PRNV), \sys~has different sensitivity to the graph density.}
\begin{table}[t]
\centering
\footnotesize
\scalebox{1}{
\begin{tabular}{ccccccc}
\hline
\multirow{2}{*}{\textbf{GraphName}} & \multicolumn{3}{c}{\textbf{RWNV}}                & \multicolumn{3}{c}{\textbf{PRNV}}                \\ \cline{2-7} 
& \textbf{SOGW} & \textbf{SGSC} & \textbf{GraSorw} & \textbf{SOGW} & \textbf{SGSC} & \textbf{GraSorw} \\ \hline
\textbf{CirculantG}                 & 1696          & 772           & \textbf{280}     & 47            & 46            & \textbf{20}      \\ \cline{1-7} 
\textbf{RandomG}                    & 10200         & 9790          & \textbf{2132}    & 304           & 290           & \textbf{64}      \\ \cline{1-7} 
\textbf{BASF}                       & 10118         & 10764         & \textbf{2171}    & 341           & 202           & \textbf{69}      \\ \hline \hline
\textbf{RandomG1}                   & 21195         & 22490         & \textbf{4083}    & 1195          & 1160          & \textbf{140}     \\ \cline{1-7} 
\textbf{RandomG2}                   & 2699          & 2705          & \textbf{670}     & 136           & 132           & \textbf{17}      \\ \cline{1-7} 
\textbf{RandomG3}                   & 544.1         & 466.3         & \textbf{201}     & 16.39         & 14.67         & \textbf{1.86}    \\ \cline{1-7} 
\textbf{RandomG4}                   & 111           & \textbf{101}  & 152.6            & 1.76          & 1.64          & \textbf{0.63}    \\ \cline{1-7} 
\textbf{RandomG5}                   & 66            & \textbf{64}   & 138.6            & 1.13          & 1.02          & \textbf{0.43}    \\ \hline \hline
\textbf{SBM1}                       & 110           & \textbf{96}   & 358              & 1.78          & 1.69          & \textbf{1.09}    \\ \cline{1-7} 
\textbf{SBM2}                       & 223           & \textbf{203}  & 633              & 3.63          & 3.45          & \textbf{2.03}    \\ \cline{1-7} 
\textbf{SBM3}                       & 179           & \textbf{165}  & 908              & 3.02          & 2.90          & \textbf{2.75}    \\ \hline
\end{tabular}}
\caption{{The wall time (in seconds) of three systems on different graphs.}}
\label{tab:syngraph-table}
\end{table}

{For RWNV, \sys~is sensitive to the density of graphs. For example, when processing the random graph whose average degree is 5, \sys~achieves 5.19$\times$ speedup compared to SOGW, but for a denser random graph whose average degree is 500, the speedup is reduced to 2.7$\times$.
When the average degree keeps increasing to 5000, \sys~is slightly slower than both SOGW and \cache. For the densest graph RandomG5, the speedup comes to 0.47$\times$. The root cause of the inefficiency of \sys~on very dense graphs is the decrease of the number of total random walks. As the density increases from RandomG1 to RandomG5, the number of total random walks decreases from 1 billion to 224 thousands in our settings. The decrease of total random walks leads to the dramatic decrease of both execution and vertex I/O time in SOGW and \cache. However, the block I/O time for \sys~remains unchanged for different densities\footnote{The I/O time of \sys~is made up with block I/Os only.}. This is because the block size and the number of blocks are similar for different graphs with different densities, and the length of the random walks is the same (80 for each walk) as well. In other words, only the execution time in \sys~decreases when the density increases, while in SOGW and \cache~, both the execution time and I/O time decreases.} 

\begin{table*}[]
\small
\centering
\begin{tabular}{c|ccc|ccc|ccc}
\hline
\multirow{2}{*}{\textbf{Dataset}} & \multicolumn{3}{c|}{\textbf{GraphWalker}}                                                                                                                                                                          & \multicolumn{3}{c|}{\textbf{GraSorw-No-LBL}}                                                                                                                                                                       & \multicolumn{3}{c}{\textbf{GraSorw}}                                                                                                                                                                               \\ \cline{2-10} 
                                  & \textbf{\begin{tabular}[c]{@{}c@{}}Wall\\ Time (s)\end{tabular}} & \textbf{\begin{tabular}[c]{@{}c@{}}Execution \\ Time (s)\end{tabular}} & \textbf{\begin{tabular}[c]{@{}c@{}}Block \\ I/O Time (s)\end{tabular}} & \textbf{\begin{tabular}[c]{@{}c@{}}Wall\\ Time (s)\end{tabular}} & \textbf{\begin{tabular}[c]{@{}c@{}}Execution \\ Time (s)\end{tabular}} & \textbf{\begin{tabular}[c]{@{}c@{}}Block \\ I/O Time (s)\end{tabular}} & \textbf{\begin{tabular}[c]{@{}c@{}}Wall\\ Time (s)\end{tabular}} & \textbf{\begin{tabular}[c]{@{}c@{}}Execution \\ Time (s)\end{tabular}} & \textbf{\begin{tabular}[c]{@{}c@{}}Block \\ I/O Time (s)\end{tabular}} \\ \hline
\textbf{LJ}                       & 137                                                              & \textbf{84}                                                            & 53                                                                     & \textbf{133}                                                     & 86                                                                     & 48                                                                     & 135                                                              & 88                                                                     & \textbf{47}                                                            \\ \hline
\textbf{TW}                       & 1366                                                             & 851                                                                    & 515                                                                    & 1399                                                             & 871                                                                    & \textbf{528}                                                           & \textbf{1302}                                                    & \textbf{793}                                                           & \textbf{509}                                                           \\ \hline
\textbf{FR}                       & \textbf{2122}                                                    & \textbf{1313}                                                          & 809                                                                    & 2200                                                             & 1362                                                                   & 838                                                                    & 2128                                                             & 1346                                                                   & \textbf{782}                                                           \\ \hline
\textbf{UK}                       & 2242                                                             & 1463                                                                   & 779                                                                    & 1867                                                             & 1189                                                                   & 677                                                                    & \textbf{1782}                                                    & \textbf{1123}                                                          & \textbf{660}                                                           \\ \hline
\end{tabular}
\caption{{Result of First-order random walk execution. We compare \sys~with GraphWalker and \sys~without learning-based block loading (\sys-No-LBL).}}
\label{tab:first-order}
\end{table*}
{Comparing to RWNV, \sys~is less density-sensitive for PRNV and the speedups are stable across different graphs. \sys~consistently outperforms SOGW and \cache~in PRNV, with a speed up from 8.8$\times$ to 2.6$\times$. The reason is that there are more walks in PRNV since we set 100 query nodes for each graph and start $4|V|$ walks from each query node. For example, even on RandomG5, the number of total random walks of PRNV is still 9 million, which is similar to that of RWNV on RandomG3. From Table~\ref{tab:syngraph-table}, we see that \sys~achieves similar speedup for RWNV on RandomG3 and PRNV on RandomG5, which is 2.7$\times$ and 2.6$\times$, respectively.}

{In summary, \sys~is able to achieve high-performance on graphs with different densities when the workload (i.e., the number of walks) is heavy; it may not be the best option for dense graphs when the workload is light.}

{\textbf{Influence of the community structure.} We use the stochastic block model\footnote{\url{https://networkx.org/documentation/stable/reference/generated/networkx.generators.community.stochastic_block_model.html}} to generate three graphs, namely SBM1, SBM2, SBM3.  
In stochastic block model, there are mainly two parameters $p$, $q$ controlling the strength of connectivity between blocks. $p$ represents the density between each two blocks, that is, the number of edges connecting the two blocks divided by the maximum possible number of edges between two blocks. $q$ represents the density inside the block, that is, the number of edges inside the block divided by the number of edges of the complete graph that vertices in the block form. A graph with larger $p$ and smaller $q$ has more edges between each block and less edges connecting vertices inside the block. For SBM1, $p=0.3$, $q=0.9$, for SBM2, $p=0.6$, $q=0.6$, and for SBM3, $p=0.9$, $q=0.3$. 
With such settings, we find that the graphs generated are extremely dense. We have tried to generate graphs with less $p$ and $q$ to make it sparser, {yet it takes too long time to generate such a graph using NetworkX.}}

{From the experimental results, we see that \cache~outperforms \sys~for RWNV tasks, while \sys~remains faster than SOGW and \cache~for PRNV tasks. The phenomena is similar to the one of graphs with different densities. As discussed before, because the three graphs are extremely dense, the performance of \sys is influenced by the workload. The detailed explanation is presented in the experiments of ``influence of different densities''.}


\subsection{{The Performance of First-order Random Walk in \sys}}
{Finally, we evaluate the performance of first-order random walk in \sys~to demonstrate the applicability of our system. We take DeepWalk as the benchmark of first-order random walk tasks, and compare \sys~with the state-of-the-art disk-based first-order random walk system GraphWalker. We also compare the results of \sys~without learning-based block loading (\sys-No-LBL), which uses the Iteration-based method mentioned in Section~\ref{subsec:revisit} to schedule the current blocks. 
The experimental results are shown in Table \ref{tab:first-order}. We see that \sys~or \sys-No-LBL achieves the best efficiency on three datasets, i.e., LJ, TW and UK, and is comparable to GraphWalker on FR. Comparing with GraphWalker, \sys~saves 21\%, 23\% and 16\% of the wall time, execution time and block I/O time on UK, respectively. With the help of learning-based block loading method, the time cost of block I/Os in \sys~is the least among all these four datasets, and it is 85\%\textasciitilde99\% of the one in GraphWalker. This is because some heavy I/Os of loading the complete block is converted to light vertex I/Os when the number of remained walks is small. In addition, comparing \sys-No-LBL with GraphWalker, both the execution time and block I/O time are similar. This demonstrates that the iterative block scheduling is effective for first-order random walks, and it is feasible to replace the state-aware block scheduling mechanism in GraphWalker with the iterative block scheduling.}

\section{Related Work}
Many systems have been designed to analyze large graphs in recent years. Some studies focused on how to migrate the benefits of distributed computing to graph processing. Pregel~\cite{Malewicz2010} proposes a synchronization model that represents various typical graph processing tasks as a series of iterations to run them on a cluster of machines. GraphLab~\cite{Low2012} proposes a model for asynchronous processing, and PowerGraph~\cite{Gonzalez2012} takes into account the power-law property of natural graphs for faster access to vertex information. However, distributed systems have high requirements of the running environment, which is expensive, and have high communication costs between nodes.

As another solution that is lightweight, inexpensive, and scalable, many single machine disk-based graph processing systems have been proposed. GraphChi~\cite{Kyrola2012} first organizes the graph data on disk in shards, thus converting random I/Os to sequential I/Os in each shard. X-Stream~\cite{Roy2013} employs a new edge-centric graph computation model that enables the system to stream the list of edges read from disk directly. GridGraph~\cite{Zhu2015} designs a more clever fine-grained subgraph partitioning to avoid loading useless information into memory and accelerate I/O processing. In addition, the features of SSD are also considered by some systems. Liu et al.~\cite{Liu2017} designed a disk-based full-granularity I/O management by reorganizing the SSD format to store graph data completely on disk, which makes the performance of dedicated SSD-based graph processing systems closer to that of memory-based graph processing system. Due to the generality, these aforementioned systems did not take into consideration the features of random walks and entail more time for random walk processing. 

Meanwhile, due to the wide applicability of random walk, several dedicated systems or frameworks have been proposed to accelerate random walk processing. Most of them are designed in-core. Shao et al.~\cite{Shao2020,Shao2021} proposed a framework for rational use of available memory, which switches between different sampling algorithms for different nodes to balance the time and space overheads. ThunderRW~\cite{Sun2021} designs a step-centric programming model to address the high CPU pipeline slots stalled due to irregular memory access in random walk tasks. UniNet~\cite{Yao2020} brings in a new edge sampler based on Metropolis-Hastings to efficiently sample the next steps of random walks and proposes a framework that provides a uniform representation of different random walk models and allows users to implement new graph representation learning models flexibly. However, these frameworks use a memory-based model and cannot provide help in the scalability of the large graph. KnightKing~\cite{Yang2019} is a distributed system aiming to optimize random walk processing and employs an efficient algorithm in second-order random walk sampling. There are also systems focusing on out-of-core random walk processing. DrunkardMob~\cite{Kyrola2013} encodes each walk and stores them in memory to support parallelism for billions of random walks. In a single block, it clusters vertices into batches and manages walks belongs to the batch together in the corresponding bucket. GraphWalker~\cite{Kyrola2013} adopts block-centric walk management and also uses a disk to store walks. It proposes state-aware block scheduling and asynchronous walk updating to reduce block I/Os. Different from these systems, \sys~ focuses on optimizing a large amount of random vertex I/Os in second-order random walk tasks, and converts these I/Os into sequential by employing bi-block execution engine and increases the I/O utilization with the help of learning-based block loading model.

\section{Conclusion}
Second-order random walk is an important method for modeling higher-order dependencies in data. The existing disk-based graph system cannot efficiently support the second-order random walk.
We proposed an I/O-efficient disk-based second-order random walk system. To reduce the massive light vertex I/Os, we developed a bi-block execution engine with a triangular bi-block scheduling strategy, which smartly converts small random I/Os into large sequential I/Os. To improve the I/O-utilization, we introduced a learning-based block loading model to select the proper block loading method automatically. Finally, we empirically evaluated our system on five large graphs, and the results demonstrated \sys~ significantly surpasses the existing disk-based random walk systems. {Furthermore, considering that the processing of second-order random walks in most of these applications is an independent phase, \sys~can be easily embedded or integrated into existing second-order random walk-based applications.}



\bibliographystyle{ACM-Reference-Format}
\bibliography{sample-base}

\clearpage
\appendix

\section{Efficiency of Different Scheduling Strategy for Current Blocks}
\label{app:scheduling}
\begin{table*}[]
    \centering
\small
    \resizebox{0.9\textwidth}{!}{%
\begin{tabular}{l|ccccc|ccccc}
\hline
\textbf{}                                                                                     & \multicolumn{5}{c|}{\textbf{LiveJournal}}                                                                                                                                                        & \multicolumn{5}{c}{Twitter}                                                                                                                                                                      \\ \hline
\textbf{}                                                                                     & Alphabet & \begin{tabular}[c]{@{}c@{}}Iteration-\\ based\end{tabular} & \begin{tabular}[c]{@{}c@{}}Min-\\ Height\end{tabular} & \begin{tabular}[c]{@{}c@{}}Max-\\ Sum\end{tabular} & GraphWalker & Alphabet & \begin{tabular}[c]{@{}c@{}}Iteration-\\ based\end{tabular} & \begin{tabular}[c]{@{}c@{}}Min-\\ Height\end{tabular} & \begin{tabular}[c]{@{}c@{}}Max-\\ Sum\end{tabular} & GraphWalker \\ \hline
\multicolumn{1}{c|}{\textbf{\begin{tabular}[c]{@{}c@{}}Number of\\ Block I/O\end{tabular}}}      & 821      & \textbf{804}                                               & 1258                                                  & 1007                                               & 963         & 924      & \textbf{919}                                               & 1296                                                  & 991                                                & 994         \\ \hline
\multicolumn{1}{c|}{\textbf{\begin{tabular}[c]{@{}c@{}}Block I/O \\ Time (Sec)\end{tabular}}} & 4.10     & \textbf{4.07}                                              & 5.72                                                  & 4.81                                               & 4.66        & 112      & \textbf{102}                                               & 146                                                   & 163                                                & 111         \\ \hline
\textbf{}                                                                                     & \multicolumn{5}{c|}{\textbf{Frindster}}                                                                                                                                                          & \multicolumn{5}{c}{\textbf{UK200705}}                                                                                                                                                            \\ \hline
\textbf{}                                                                                     & Alphabet & \begin{tabular}[c]{@{}c@{}}Iteration-\\ based\end{tabular} & \begin{tabular}[c]{@{}c@{}}Min-\\ Height\end{tabular} & \begin{tabular}[c]{@{}c@{}}Max-\\ Sum\end{tabular} & GraphWalker & Alphabet & \begin{tabular}[c]{@{}c@{}}Iteration-\\ based\end{tabular} & \begin{tabular}[c]{@{}c@{}}Min-\\ Height\end{tabular} & \begin{tabular}[c]{@{}c@{}}Max-\\ Sum\end{tabular} & GraphWalker \\ \hline
\multicolumn{1}{c|}{\textbf{\begin{tabular}[c]{@{}c@{}}Number of\\ Block I/O\end{tabular}}}      & 1430     & 1408                                                       & 2081                                                  & \textbf{1399}                                      & 1410        & 1099     & \textbf{1088}                                              & 1909                                                  & 1645                                               & 1561        \\ \hline
\multicolumn{1}{c|}{\textbf{\begin{tabular}[c]{@{}c@{}}Block I/O \\ Time (Sec)\end{tabular}}} & 167      & 157                                                        & 225                                                   & \textbf{155}                                       & 156         & 251      & \textbf{235}                                               & 406                                                   & 351                                                & 335         \\ \hline
\end{tabular}%
}
    \caption{{The performance of block I/O with different current block scheduling strategy. The number of blocks for each graph is consistent with Table \ref{tab:datasets}.} }
    \label{tab:minimal-block-io}
\end{table*}
{The minimal current block I/Os problem is NP-hard, and it is interesting to know the effectiveness guarantee of the iteration-based method. In general, we should compare the iteration-based method to the ground truth or the optimal solution. 
However, for the minimal current block I/Os problem, it is difficult (or infeasible) to compute the ground truth even on small datasets. For a set of $k$ sequences with length $n$, the space and time complexity to compute their accurate SCS is $\mathcal{O}(n^k)$ using dynamic programming.
That is to say, the time and space complexity is $\mathcal{O}(n^k)$ when computing the accurate ground truth of the minimal current block I/Os for $k$ walks with $n$ steps. On a tiny graph with tens of nodes, a simple random walk task generates hundreds of walks, and each walk moves tens of steps. 
For example, for a simple tasks with 100 walks (10 steps for each walk), the computation of the ground truth requires 4 googol ($4\times10^{100}$) bytes of memory, and takes billions of years to compute such an answer. 
Apparently, it is not feasible to perform such a task on general commodity machines within finite time. Similarly, P. Barone et.al.~\cite{Barone2001} states that the computation of the ground truth of such a task is not acceptable due to time and space constraints when $k$ reaches 5\textasciitilde8.
}

{In the field of SCS problems in which it is not possible to compute the exact ground truth, a comparison between the proposed method and other existing method is presented~\cite{Ning2005,Ning2006}. Similarly, we compare our iteration-based method with several other well-known heuristic solutions for SCS problems. Note that, as discussed in Section~\ref{subsec:revisit}, not all approximated solution for SCS is feasible in our situation. This is because the minimal current block I/O problem defines an \textit{online} SCS problem, that is, the next item of all sequences is not aware unless the current item in the supersequence is chosen. In the context of \sys, only the historical blocks are aware, thus, the global information of the sequences is missing when scheduling the next current block.
Therefore, we only compare the existing solutions that are feasible for online SCS problems, and briefly make a list of how they are used in random walk tasks as follows:}
\begin{itemize}
    \item \textit{The Alphabet algorithm.} The current block is scheduled from $b_0$ to $b_{N_B}$. The scheduling starts from $b_0$ again if there are walks not terminated. The approximation ratio of this method is $N_B$~\cite{Ning2006}.
    \item \textit{Our Iteration-based method.} It is similar to the Alphabet algorithm. The only difference is that when there is no walk in a certain block, it will be skipped and not chosen as the current block. The approximation ratio of this method is also $N_B$.
    \item \textit{The Min-Height method.} The block currently containing the walk with shortest steps is chosen as the current block.
    \item \textit{The Max-Sum method.} The block currently containing most walks is chosen as the current block.
    \item \textit{GraphWalker method.} This is a mix of the Max-Sum and Min-Height method. With a probability $p$ the current block is chosen by the Max-Sum method, and $1-p$ chosen by the Min-Height method. In the experiment, we follow the settings in GraphWalker and set $p$ to 0.8.
\end{itemize}
{Note that the last three methods do not have an approximation ratio~\cite{Ning2006}.}

{
Focusing on the scheduling of current blocks, there is no difference between higher-order random walk models and first-order random walk models. 
Therefore, for simplicity, we perform DeepWalk on four datasets, and start 10 walks from each node by setting the walk length to 80. The results are shown in Table \ref{tab:minimal-block-io}. 
We can see that the Iteration-based method incurs the fewest block I/O and achieves the best performance in block I/O time for three datasets, LiveJournal, Twitter and UK200705. On Friendster, although the Max-Sum method achieves the best result for both block I/O times and total time of block I/O, the Iteration-based method achieves the second-best result, and the difference between these two methods is very small. On the other hand, the performance of a method may vary widely on different datasets. For example, the Max-Sum method achieves the best result on Friendster, yet on UK200705 it performs poorly, which incurs 51\% more block I/Os comparing with the Iteration-based method, which achieves the best performance on this dataset. 
In conclusion, among the existing heuristic online solutions, no single method performs optimally on all datasets, and the performance of the same method on different datasets may vary widely.
But in general, the Iteration-based method achieves the best result in most cases.}

\section{Correctness Proof of Algorithm 1} \label{app:correctness}

{Given a termination condition $T$ which can be reached for each walk in finite steps, let $N$ be the total number of walks, $F$ be the set of walks that reach $T$, $U$ be the set of walks that have not reached $T$, then we have $N=|F| + |U|$. 
\begin{enumerate}
\item In the beginning, $|U| = N$. Since we ensure that the current vertex and the previous vertex of a walk belongs to different blocks after the initialization, it means no walk $w_u^v$ such that $B(u) = B(v)$ exists. During the initialization, a walk either reaches $T$ and is added into $F$ or walks out of the $B(s)$, where $s$ represents its source vertex. Therefore, we still have $|F| + |U| = N$ after the initialization stage. 
\item In each time slot, no walk $w_u^v$ such that $B(u) = B(v)$ exists (ensured by Line 22 in Algorithm 2). Therefore, for a walk $w_u^v$, either $B(u) > B(v)$ or $B(v) > B(u)$. With skewed walk storage, walks such that $B(u) > B(v)$ is associated with $B(v)$. Since $B(u) > B(v)$, from triangular block scheduling method we know that there exists a state such that $B(u)$ is loaded as the ancillary block and $B(v)$ is the current block. Consequently, the walk $w_u^v$ gets updated by at least one step forward. We can get the same results for walks such that $B(v) > B(u)$. Therefore, all walks get updated by at least one step forward in one time slot, and either reach $T$ and are added into $F$ or still stay in $U$. So after each time slot, $|F| + |U| = N$ still holds. 
\item The algorithm terminates when $|U| = 0$. Since in each time slot all walks get updated by at least one step forward, they can reach $T$ in finite steps. Let $s_i$ be the number of steps of the $i$-th walk when it terminates and $S = max(s_i)$, then after $S$ time slots, $|F| = N$. 
We know that $|F| + |U| = N$ holds after each time slot, and when $|F| = N$, we have $|U| = 0$. Therefore, the termination condition (i.e., $|U| = 0$) of the algorithm can be reached, and the algorithm is correct. 
\end{enumerate}}

\section{Complexity Analysis of Algorithm 1} \label{app:complexity}

{Algorithm \ref{alg:triangular-scheduling} is made up with three components: 1. block loading, which includes the loading of current blocks and ancillary blocks; 2. walk processing, which includes the loading and execution of walks; 3. walk management, which collects walks to buckets. In the following, we discuss about the space and time cost one-by-one.}

{\textbf{Cost of block loading.} Let $t_l$ be the time of loading one block into memory, $s_b$ be the bytes that a block occupies, {$L$ be the length of each walk}, and $N$ the number of total walks. In each time slot, all walks move forward by at least one step. Therefore, the total number of time slot is at most $L$. From equation (3) we know that in each time slot, there are at most $\frac{1}{2}(N_B + 2)(N_B - 1)$ block I/Os. 
Therefore, the time cost of block loading is at most $\frac{1}{2}(N_B + 2)(N_B - 1)L t_l$.
Since there are at most two blocks in memory at the same time, the space cost is $2s_b$.}

{\textbf{{Cost of walk processing.}} Let $t_s$ be the average time cost to sample one step, $t_w$ be the average time to load a walk, and $s_w$ be the bytes that a walk occupies. In each time slot, all walks are loaded into memory, and the walk loading time is $NLt_w$. The total sampling time is $NLt_s$. For each current block, the average number of walks associated with it is $N/N_B$, and so the memory required for storing walks in memory is $s_wN/N_B$. }

{\textbf{Cost of walk management.} The array of storing the walks in memory is called the \textit{current walk array}. Let $t_b$ be the time of collecting one walk into its belonging bucket. Since a walk is updated at least one step forward after it has been collected into its belonging bucket, a walk is collected to bucket by at most $L$ times. Then for $N$ walks, the total collecting time cost is at most $NLt_b$. Note that after the collection of each walk, the walk is removed from the current walk array, and thus the buckets bring in no extra space.}

{Summing these costs up, the time complexity of Algorithm 1 is $\mathcal{O}(N_B^2Lt_l + NL(t_s + t_b))$, and the space complexity is $\mathcal{O}(2s_b + s_wN/N_B)$.}

\end{document}